\documentclass[11pt,a4paper]{article}
\pdfoutput=1                  
\usepackage[pdftex]{graphics} 
 
\usepackage{amsmath,amsthm,amssymb}
\usepackage{graphics,graphicx}
\usepackage{epsfig}
\usepackage{multicol}
\usepackage{color}
\usepackage{ifpdf}
\makeatletter
\@addtoreset{equation}{section}
\makeatother

\begin{document}

\begin{center}

\Large{\bf A Generalisation of the Nielsen-Olesen Vortex: Non-cylindrical Strings in a modified Abelian-Higgs Model}

\end{center}

\begin{center}

\large{\bf Matthew Lake,} ${}^{a, b,}$\footnote{m.lake@qmul.ac.uk} \large{\bf and John Ward} ${}^{c,}$\footnote{jwa@uvic.ca}

\end{center}

\begin{center}

\emph{ ${}^a$ Center for Research in String Theory, Queen Mary, University of London \\ Mile End Road, London E1 4NS, UK \\}

\vspace{0.1cm}

\emph{ ${}^{b}$ Astronomy Unit, School of Mathematical Sciences, Queen Mary, University of London \\ Mile End Road, London E1 4NS, UK \\}

\vspace{0.1cm}

\emph{${}^{c}$ Department of Physics and Astronomy, University of Victoria, Victoria, BC \\ V8P 1A1, Canada }

\end{center}
\begin{abstract}
We modify the standard Abelian-Higgs model by introducing spatially-dependent couplings for the scalar and vector fields. 
We investigate static, non-cylindrically symmetric solutions of the resulting field equations and propose a pinch solution 
which interpolates between degenerate vacua along the string, labelled by $\pm \left|n\right|$. 
This configuration corresponds to a vortex which shrinks to the Planck scale before re-emerging as an anti-vortex, 
resulting in the formation of a bead pair with one bead either side of the intersection. 
The solution is then topologically stable. 
A key assumption is that quantities such as phase and winding number, along with those which depend on them like the magnetic flux, 
become undefined at the Planck scale so that regions of opposite winding may be joined via a Planck-sized segment of neutral string. 
Similarities between this solution and the extra-dimensional windings of strings in type IIB string theory are discussed 
and a correspondence between field theory and string theory parameters is suggested. 
The spatial-dependence of the field couplings is found to have a natural interpretation in the string picture and 
results from the variation of the winding radius, giving rise to a varying (effective) string coupling. 
An interesting result is an estimate of the Higgs mass (at critical coupling) in terms of the parameters which define the Klebanov-Strassler 
geometry and which, in principle, may be constrained by cosmological observations.
\end{abstract}

\section{Introduction}

Previous studies of strings in backgrounds with compact extra-dimensions have led to an increased understanding 
of the formation and evolution of string loops with non-trivial windings in the internal space.  
Both generic scenarios and compactification schemes specific to type IIB string theory have been 
investigated \cite{Avgoustidis:2004zt}-\cite{LakeThomasWard:JCAP2010}.
\\
\indent
In the case of static string loops, it has been shown that the 
presence of a lifting potential in the compact space traps the windings - giving rise to 
loops with step-like winding configurations referred to in the literature as cosmic necklaces \cite{Matsuda:2004bg}-\cite{Matsuda:2005fb}.
From a four-dimensional perspective the windings appear as a series of monopoles or ``beads" connected by ordinary sections of string. 
Superficially these resemble the standard string-monopole networks found in field-theoretic models \cite{Vilenkin+Shellard:book} 
but their behaviour is, in many ways, fundamentally different. 
In stark contrast to previous predictions based on field theory defects, it was found that the gravitational collapse of 
necklaces formed in the class of backgrounds defined by the Klebanov-Strassler (KS) geometry \cite{Klebanov:2000hb} 
leads to the formation of primordial black holes (PBHs) \cite{Carr:2009jm} during a window in the early universe,
followed by the formation of potential dark matter (DM) relics in the scaling regime \cite{LakeThomasWard:JHEP2009}. 
This is almost the complete reverse of the standard predictions for string-monopole networks \cite{Berezinsky:1997td}.
\\
\indent
The root cause of this difference appears to be a time-dependent bead mass in the necklace model, as opposed to the 
constant bead mass of true monopoles connected by strings. 
This arises from the time-dependence of the lifting potential and is a somewhat unexpected result. 
Initial investigations of string necklaces assumed the existence of a constant potential and hence a constant bead mass, 
though these were based solely on generic arguments \cite{Matsuda:2004bg}-\cite{Matsuda:2005fb}. 
The work presented in \cite{LakeThomasWard:JHEP2009} represents the first explicit realisation of necklace formation in string theory, 
in which it was shown that these assumptions must be modified, at least for the class of backgrounds considered.
This raises two interesting possibilities: Either the formation of necklace-like objects is possible only in string theory, 
or there exist previously unknown solutions in field theory models which are equivalent to the objects described by strings with step-like windings.
\\
\indent
At present there is no known field theory analogue of necklaces formed from extra-dimensional windings. 
In particular, there is no known way to produce string-monopole networks with time varying bead mass. 
The question of whether a corresponding field theory model exists is important because, 
if the formation of necklace-type objects is specific to models with compact extra dimensions, their predicted impact on observable parameters 
could be used to obtain experimental evidence in favour of string theory, or at least in favour of higher-dimensional models. 
The aim of this paper is to investigate the possibility of such a configuration. 
We seek to establish a relation between the topological winding number, $(n)$, of a defect string (in four spacetime dimensions) 
and the physical winding number, $(n_w)$, of an $F$/$D$-string in a higher-dimensional background. 
We also seek to establish a field-theoretic model in which ``bead"/``pinch" formation occurs dynamically, by analogy with the wound-string case \cite{Avgoustidis:2004zt}-\cite{LakeThomasWard:JCAP2010}. 
\\
\indent
For simplicity we consider the most basic field-theoretic model of local string formation - the Abelian-Higgs model.
A non-cylindrical, or ``pinched" string solution - which represents generalisation of the well known Nielsen-Olesen vortex \cite{Nielsen_Olesen} - 
naturally arises if the usual Abelian-Higgs action is modified to include 
coordinate-dependent couplings in both the scalar and vector fields (i.e. $\sqrt{\lambda}^{eff}(z)$ and $e^{eff}(z)$ in our gauge choice, where $z$ represents the position along the length of the string). This appears to mimic the behaviour of the type of necklaces discussed above.
\\
\indent
We propose the pinched string solution as a model corresponding to wound-string necklaces, and argue that a time-dependent bead mass may be obtained.
The main basis for the proposed correspondence is an analysis of the four-dimensional effective tension of the wound-strings, 
together with the periodically varying tension of the pinched string, which take the same form for appropriate ansatz choices. 
Following the correspondence between string theory and field theory parameters suggested by this comparison, we propose
a geometric interpretation of the field-theoretic terms, such as gauge flux and topological winding number, in the string picture. 
Although the introduction of $z$-dependence in the original field couplings may appear unnatural, it has a very 
natural interpretation in the string picture due to the relation between the three-sphere radius $R$ and the string coupling $g_s$ ($R^2 \sim b_0Mg_s\alpha'$).
\\
\indent
As the interchange between vortices and anti-vortices in the field-theoretic model necessarily involves the consideration of 
sub-Planckian scales, we present a hypothetical model for discretising the Planck-scale structure of the vortex. 
Though such discretisation is necessary to prevent divergences in the Euler-Lagrange equations, it is not intended to be literal.
Whatever the exact nature of the physical limit imposed by the Planck scale, it is likely to determine a limit on the process of measurement 
rather than implying the outright discretisation of spacetime. 
We adopt this method only as an approximation to an (as yet) unknown theory of the quantum structure of fields on the smallest scales, and 
present an argument for its validity (as a toy model) based on matching boundary solutions of Planck-sized regions to well known solutions 
valid on scales $\Delta r, \ \Delta z \geq l_p$. Once again, we find that the localisation of the (classical) 
field-theoretic string core on scales $r \geq \mathcal{O}(l_p)$ admits a very natural interpretation in terms of the 
quantum constraints on the corresponding wound $F$/$D$-strings.
\\
\indent
The structure of this paper is as follows: In section 2 we present a brief overview of the Abelian-Higgs model, 
including the general (covariant) form of the Euler-Lagrange equations. 
For completeness, and to allow easy comparison with later calculations of the $z$-dependent tension of the pinched string $\mu_{|n|}(z)$, 
an outline of the calculation of the constant string tension $\mu_{|n|}$ for the cylindrically symmetric ansatz is reviewed in section 3.
In section 4 we introduce a non-cylindrically symmetric ansatz based on a specific discretisation scheme 
in the Planck-scale region of the vortex core (in which vorticity itself may no longer be defined), and introduce the $z$-dependent 
couplings $\sqrt{\lambda}^{eff}(z)$ and $e^{eff}(z)$. 
From this we derive the specific form of the equations of motion (EOM) to verify that the pinched string configuration is a valid solution. 
Under a set of reasonable assumptions, these are found to reduce to simplified forms which (for fixed $z$) are structurally 
equivalent to the EOM for the cylindrically symmetric case. 
This allows analogous asymptotic and small $r$ solutions for the scalar and gauge fields 
but their behaviour is now, generally, $z$-dependent. 
We show that the $z$-dependent tension depends crucially on the form of a generic, dimensionless, periodic function $G(z)$ which varies between zero 
and one but which, to good approximation, is independent of the physics of the vortices in the Planck-scale regions.
We then discuss the effective four-dimensional tension of an $F$/$D$-string with linear winding ansatz 
at the tip of the KS throat, and compare with the results for our pinched string. 
For an appropriate and natural choice of $G(z)$ we see that this enables a correspondence between the field theory and the string theory parameters to be inferred. 
This section also contains a more detailed analysis of the relation between the scalar and vector fields in the Higgs model, 
and their geometric interpretation in the theory of wound strings. 
\section{Revisiting the Abelian-Higgs model}

Using the $(+---)$
metric convention, and introducing the covariant derivative $D_{\mu} = \partial_{\mu} + ie A_{\mu}$ (together with its conjugate), 
the covariant form of the Abelian-Higgs action is \cite{Hindmarsh:1994re}
 \begin{equation} \label{eq:action1}
S = \int d^4x \sqrt{-g} \left(D_{\mu}\phi g^{\mu \nu}\overline{D}_{\nu}\overline{\phi} - \frac{1}{4} F_{\mu \nu} F^{\mu \nu}
-\frac{1}{4}\lambda(\phi \overline{\phi} - \eta^2)^2\right).
\end{equation}
Varying this action with respect to the various fields yields the resultant EOM, 
assuming that the gauge field and scalar derivatives vanish at the boundary,
\begin{eqnarray} \label{eq:ELeqn1}
0 &=&\frac{1}{\sqrt{-g}}D_{\mu}\left(\sqrt{-g} g^{\mu \nu} D_{\nu}\phi \right) - \frac{\lambda}{2}\phi(\phi\overline{\phi} - \eta^2) \nonumber \\
0 &=& \frac{1}{\sqrt{-g}}\overline{D}_{\mu}\left(\sqrt{-g} g^{\mu \nu} \overline{D}_{\nu}\overline{\phi} \right) - \frac{\lambda}{2}\overline{\phi}
(\overline{\phi}\phi - \eta^2),
\end{eqnarray}
with the corresponding Maxwell equation now taking the form
\begin{equation} \label{eq:current1}
\frac{1}{\sqrt{-g}}\partial_{\mu}\left(\sqrt{-g}F^{\mu \nu}\right) = j^{\nu}.
\end{equation}
The $U(1)$ current is 
\begin{equation} \label{eq:current2} 
j^{\nu} = -ie g^{\mu \nu}\left(\overline{\phi} D_{\mu}\phi - \phi \overline{D}_{\mu}\overline{\phi} \right).
\end{equation}
The static, cylindrically symmetric ansatz then takes the form
\begin{eqnarray} \label{eq:scs_ansatz1}
\phi_{n}\left(R_s,\theta\right) &=& \eta f(R_s) e^{in\theta} \nonumber\\
A_{n\theta}(R_v) &=& -\frac{n}{e}\alpha(R_v)
\end{eqnarray}
where
\begin{equation} \label{eq:R_sR_v}
R_s = \frac{r}{r_s}, \hspace{0.5cm} R_v = \frac{r}{r_v}
\end{equation}
are dimensionless variables and $r_s, r_v$ are the length scales of the scalar and vector cores, respectively. 
These are fixed by the Compton wavelengths of the associated scalar and vector bosons and, in the case of the vector core, by the winding number number of the vortex $|n|$ (which is also equal to the number of units of magnetic flux), to be \cite{Bogomolnyi:1976_1}
\begin{eqnarray} \label{eq:compton1}
r_s &=& (m_s)^{-1} \approx (\sqrt{\lambda}\eta)^{-1} \nonumber\\
r_v &=& \sqrt{|n|}(m_v)^{-1} \approx \sqrt{|n|}\left(2 e \eta\right)^{-1},
\end{eqnarray}
Here $f(R_s)$ and $\alpha(R_v)$ are dimensionless real functions satisfying the conditions
\begin{equation} \label{eq:f_bc}
f(R_s) = \left \lbrace
\begin{array}{rl}
0 & \text{if } R_s=0 \ \ \ (r=0) \\
1 & \text{if } R_s \rightarrow \infty \ (r >> r_s)
\end{array}\right.
\end{equation}
and
\begin{equation}\label{eq:alpha_bc}
\alpha(R_v) = \left \lbrace
\begin{array}{rl}
0 & \text{if } R_v=0 \ \ \ (r=0) \\
1 & \text{if } R_v \rightarrow \infty \ (r >> r_v).
\end{array}\right.
\end{equation}
We can then define the scalar and gauge field equations by
\begin{equation} \label{eq:scs_eom1} 
0 = \frac{d^2 f}{d R_s^2}  + \frac{1}{R_s}\frac{d f}{d R_s} 
+ \frac{n^2 f}{R_s^2}(\alpha^2-1)-\frac{f(f^2-1)}{2}
\end{equation}
\begin{equation} \label{eq:scs_eom2}
0 = \frac{d ^2 \alpha}{d R_v^2} - \frac{1}{R_v}\frac{d \alpha}{d R_v} - |n|f^2(\alpha-1).
\end{equation}
Here we have manipulated the original forms of the EOM (which come from simply substituting the ansatz (\ref{eq:scs_ansatz1}) 
into (\ref{eq:ELeqn1})-(\ref{eq:current1}))by multiplying the original scalar equation by $r_s^2$ to get (\ref{eq:scs_eom1}) 
and the original vector equation by $r^2 r_v^2$ in order to get (\ref{eq:scs_eom2}). 
This allows us to define both the scalar and vector EOM in terms of the dimensionless variables $R_s$ and $R_v$.
\\
\indent
Although multiplying the original form of our equations by powers of $r$ may be hazardous in the asymptotic limit, 
multiplying the vector EOM through by a factor of $r^2$ causes no problems for $r \rightarrow \infty$ as each term in the 
equation still goes to zero independently 
as $R_s,R_v \rightarrow \infty$ and $f(R_s), \alpha(R_v) \rightarrow 1$.
\\
\indent
However, defining the parameter 
\begin{equation} \label{eq:beta}
\beta = \left(\frac{r_v}{r_s}\right)^2
\end{equation}
we can rewrite the scalar equation as
\begin{equation} \label{eq:scs_eom1_mod}
0 = \frac{d^2 f}{d R_v^2}   + \frac{1}{R_v}\frac{d f}{d R_v} + \frac{n^2 f}{R_v^2}(\alpha^2 - 1) -\frac{1}{2}\beta f(f^2-1).
\end{equation}
Treating the ratio $\beta$ as a numerical constant then allows us to rewrite both the scalar and vector EOM 
in terms of a single dimensionless variable $R_v$, and adopting the form (\ref{eq:scs_eom1_mod}) for the scalar EOM 
is equivalent to assuming $f=f(R_v)$ instead of $f=f(R_s)$ in the ansatz (\ref{eq:scs_ansatz1}). 
This may appear counter-intuitive because we expect the length scale $r_s$ to determine the width of the scalar core 
(i.e. the region over which $f \approx 0 \rightarrow f \approx 1$). However, the fact that (\ref{eq:scs_eom1}) 
and (\ref{eq:scs_eom1_mod}) are algebraically equivalent shows that we may assume either functional form for $f$ in our initial ansatz. 
Both equations have the same approximate solutions in the asymptotic and small $r$ limits and 
it is the value of $r_s$ which characterises the small $r$ solution, though both scales $r_s$ and $r_v$ play a role in the asymptotics 
(at least when the EOM are solved as a coupled pair \cite{perivalopolous:1993}). 
\\
\indent
We find it convenient to use the form (\ref{eq:scs_eom1_mod}) instead of (\ref{eq:scs_eom1}) for the scalar EOM, though ``large" and ``small" $r$ for both forms of the equation must still be defined with respect to $r_s$ rather than $r_v$ as for the vector EOM (\ref{eq:scs_eom2}). Before concluding this section we note the approximate solutions for the functions $f$ and $\alpha$ within the scalar and vector cores when the EOM are solved as an uncoupled pair, i.e.
\begin{equation} \label{eq:small_r2}
f(R_s) \sim R_s^{\left|n\right|},
\end{equation}
\begin{equation} \label{eq:small_r2a}
\alpha(R_v) \sim R_v^2.
\end{equation}
Asymptotic solutions to the uncoupled equations may be found in the usual literatue (c.f. \cite{Vilenkin+Shellard:book,Hindmarsh:1994re,Preskill}) and solutions valid for both small and large $r$ when the EOM are solved as a coupled pair may be found in \cite{perivalopolous:1993,LakeThesis}. To lowest order, the small $r$ solutions for both functions are not affected by the coupling, so that the expressions above remain approximately valid. 
\section{Calculation of the (constant) string tension for a cylindrically symmetric string} 
The formula for the tension of an Abelian-Higgs vortex string is
\begin{eqnarray} \label{eq:Mu_1a}
\mu = \int r dr d\theta \left\{D_{\mu}\phi \overline{D}^{\mu}\overline{\phi} + \frac{1}{2}(\vec{E}^2 + \vec{B}^2) + V(\left|\phi\right|)\right\}
\end{eqnarray} 
which gives 
\begin{equation} \label{eq:mu_1}
\mu \sim 2\pi \eta^2 
\end{equation}
at critical coupling ($\beta=1$) and
\begin{equation} \label{eq:mu_2}
\mu \sim 2\pi \eta^2 \ln\left(\sqrt{\beta}\right)
\end{equation}
when $\beta>1$ (for $|n|\sim \mathcal{O}(1)$) \cite{Vilenkin+Shellard:book,Preskill}. This result may be verified explicitly by splitting the integral (\ref{eq:Mu_1a}) into three parts and using the appropriate approximate solutions. Within the scalar core region, $0 \leq r \leq r_s$, we may approximate the functions $f$ and $\alpha$ by $f \sim \frac{r^{\left|n\right|}}{r_s^{\left|n\right|}}$ and $\alpha \sim \frac{r^2}{r_v^2}$. In the region $r_s \leq r \leq r_v$ we assume that $\alpha \sim \frac{r^2}{r_v^2}$ and $f \sim 1$ and on scales $r>r_v$ we may assume that $f,\alpha \sim 1$ so 
that $D_{\mu}\phi \overline{D}^{\mu}\overline{\phi} \sim \left(\frac{\partial f}{\partial r}\right)^2 + \frac{f^2\left|\alpha^2-1\right|}{r^2} \rightarrow 0$ 
and the gauge field contribution effectively cancels the energy density contribution of the scalar field gradient. 
The overall contribution to the energy density of the covariant gradient term is therefore:
\begin{eqnarray} \label{eq:calc_16}
\int_{0}^{2\pi}d\theta \int_{0}^{\infty} D_{\mu}\phi \overline{D}^{\mu}\overline{\phi} r dr \approx 2\pi \eta^2 \left|n\right| + 2\pi \eta^2 \left|n\right|^2 
\ln\left(\sqrt{\beta}\right)
\end{eqnarray} 
Turning our attention to the gauge field term, we see that $\vec{E}=0$ and the only non-zero component of $\vec{B}$ is 
\begin{eqnarray}
B_z = F_{r\theta} = \partial_r A_{\theta} 
= -\frac{n}{e}\frac{d\alpha}{dr},
\end{eqnarray} 
so that
\begin{eqnarray}
\vec{B}^2 = F_{r\theta}F^{r\theta} = \frac{n^2}{e^2 r^2}\left(\frac{d\alpha}{dr}\right)^2
\end{eqnarray} 
Integrating over the vector core, $0 \leq r \leq r_v$, for which $\alpha \sim \frac{r^2}{r_v^2}$ then gives
\begin{eqnarray}
\int_{0}^{2\pi}d\theta \int_{0}^{r_v} \frac{1}{2}\vec{B}^2 r dr \approx 2\pi \eta^2 \left|n\right|
\end{eqnarray}
where we have used the definition $r_v \approx \sqrt{|n|}(\sqrt{2}e\eta)^{-1}$. 
For $r>r_v$ we may take $\alpha \approx 1$ so that $\vec{B}^2 \propto \left(\frac{d\alpha}{dr}\right)^2 \rightarrow 0$. 
The leading order contribution of the potential term is 
\begin{eqnarray}
\int_{0}^{2\pi}d\theta \int_{0}^{r_s} V(\left|\phi\right|)r dr \approx  \frac{\pi \eta^2}{\left|n\right|+2}
\end{eqnarray}
and, summing all contributions, the total tension becomes
\begin{equation}
\mu_{|n|} \sim 2 \pi \eta^2 |n| \left(2+\ln \left( \frac{r_v}{r_s}\right)|n| + \ldots \right).
\end{equation}  
Although, for $\beta>1$, this expression is dominated by the terms proportional to $\left|n\right|^2$ at large winding, at critical coupling $(\beta=1)$ we may approximate the tension by
\begin{eqnarray} \label{eq:convenient_mu1}
\mu \sim 4\pi \eta^2 \left|n\right|.
\end{eqnarray}   
Note that the numerical factor containing $\pi$ is to some degree arbitrary and depends on the definition of the vector core radius (here we have employed the definition $r_v \sim \sqrt{|n|}(\sqrt{2} e \eta)^{-1}$). The key points are that the leading order contribution at critical coupling is proportional to $|n|$, which agrees with the expected result, and that at non-critical coupling the tension is logarithmically dependent on $\beta$. Note that the inclusion of the factor of $\sqrt{|n|}$ in the expression for $r_v$ ensures the first of these conditions and this, in turn, ensures the stability of the vortex according to the famous result by Bogomol'nyi \cite{Bogomolnyi:1976_1,Bogomolnyi:1976_2} (c.f. also \cite{de_Vega:1976_1}).
\\
\indent
In the following section we introduce a non-cylindrically symmetric ansatz for the pinched string
and solve the resulting EOM. We show that under a set of reasonable physical assumptions, these equations 
take on analogous forms to those in the more familiar cylindrical case, 
but with the substitutions $r_s \rightarrow r_s^{eff}(z)$, $r_v \rightarrow r_v^{eff}(z)$. 
In other words, the structure of the vortices remains essentially the same except that the radii of the scalar and vector cores become functions 
of their position along the string. Therefore the calculation of the tension (for fixed $z$)
remains the same as that outlined above, apart from the introduction of a new radial magnetic flux term $\sim F_{z\theta}F^{z\theta}$ 
(associated with the $z$-dependent variation of $A_{\theta}$) and of a new $z$-derivative component in the 
scalar gradient which give rise to an additional periodic term.
Although the resulting tension $\mu_{|n|}(z)$ is periodic in $z$, it still carries a leading order 
constant term $\sim 2\pi \eta^2 \left|n\right|$, derived in an analogous way to that in (\ref{eq:convenient_mu1}), which may be identified with a similar leading order constant term $\sim 2\pi a_0 T_{1}$ in the expression for the effective four-dimensional tension of a wound $F/D$-string. 
Additionally, by analogy with $\beta = (r_v/r_s)^2$, we may define the parameter $\beta^{eff}(z) = (r_v^{eff}(z)/r_s^{eff}(z))^2$. 
Taking appropriate ansatz choices for $r_s^{eff}(z)$ and $r_v^{eff}(z)$ we have $\beta^{eff}(z) \equiv \beta = 1$ for $r_s = r_v$ 
and the resulting analogue of the logarithmic term $\ln(\sqrt{\beta})$ (i.e. $\ln(\sqrt{\beta^{eff}(z)})$) also vanishes.
\section{Introduction of the modified action and pinched string ansatz} 

We now introduce $z$-dependent couplings in both the scalar and vector sectors of the Abelian-Higgs model. 
Although the couplings are coordinate-dependent, they are not treated as fundamental fields in the same way as $\phi$ and $\vec{A}$.
\footnote{We will see that this phenomenological device admits a natural interpretation in the corresponding string picture and is related specifically to the string embedding.}
We are interested in static, non-cylindrically symmetric solutions to the resulting covariant EOM
and must therefore modify the ansatz (\ref{eq:scs_ansatz1}) to include $z$-dependence. 
Crucially we wish our new solution to represent a string which interpolates between energetically degenerate regions of vortex/anti-vortex solutions to the 
EOM and we anticipate the existence of a Planck-sized region in which vorticity itself cannot be defined and 
which separates neighbouring string sections labelled by $\pm \left|n\right|$. 
To this end we will also find it necessary to parameterise our new couplings in terms of the winding number
\begin{eqnarray} \label{eq:subs0}
\sqrt{\lambda} \rightarrow \sqrt{\lambda}_{|n|}^{eff}(z) &=& \sqrt{\lambda} G_{\sqrt{\lambda}|n|}(z) \nonumber\\
e \rightarrow e_{|n|}^{eff}(z) &=& e G_{e|n|}(z)
\end{eqnarray}
where $G_{\sqrt{\lambda}|n|}(z) \in [1,r_s/|n|l_p]$ and $G_{e|n|}(z) \in [1,r_v/|n|l_p]$ are dimensionless functions.
\\
\indent
This implies the existence of $z$-dependent scalar and vector boson masses (also paramaterised by $|n|$) 
and hence of $z$-dependent effective radii for the scalar and vector cores. In other words, a non-cylindrical string can be found from the correspondence
\begin{eqnarray} \label{eq:subs1}
r_s \rightarrow r_{s|n|}^{eff}(z) &=& (\sqrt{\lambda}_{|n|}^{eff}(z) \eta)^{-1} \nonumber\\
r_v \rightarrow r_{v|n|}^{eff}(z) &=& \sqrt{|n|}(\sqrt{2} e_{|n|}^{eff}(z) \eta)^{-1}.
\end{eqnarray} 
By analogy with the cylindrically symmetric case we may define new dimensionless variables using
\begin{eqnarray} \label{eq:subs2}
R_s \rightarrow R_{s|n|}^{eff}(z) &=& \frac{r}{r_{s|n|}^{eff}(z)} \nonumber\\
R_v \rightarrow R_{v|n|}^{eff}(z) &=& \frac{r}{r_{v|n|}^{eff}(z)}
\end{eqnarray} 
and new $r, z$-dependent functions (also parameterised by $|n|$) $F_{|n|}(r,z)$ and $A_{|n|}(r,z)$ via
\begin{eqnarray} \label{eq:subs3}
F(r) \equiv F(R_s) \rightarrow F_{|n|}(r,z) &\equiv& F(R_{s|n|}^{eff}(z)) \nonumber\\
A(r) \equiv A(R_v) \rightarrow A_{|n|}(r,z) &\equiv& A(R_{v|n|}^{eff}(z)).
\end{eqnarray}
However we leave the precise form of the $z$-dependence unspecified. In order to describe the existence of regions with $+|n|$ and $-|n|$ winding, as well as the
Planck-sized region which marks the transition (and in which $|n|$ becomes undefined), we must introduce a step-like function $H_{\left|n\right|}(z)$ into the field ansatz which takes three distinct values
\begin{eqnarray}
H_{\left|n\right|}:z \rightarrow \left\{-1,0,+1\right\}, \forall z \in \mathbb{R}.
\end{eqnarray}
Our modified ansatz then takes the form
\begin{eqnarray} \label{eq:nscs_ansatz1}
\phi_{\left|n\right|}(r,\theta,z) &=& F_{\left|n\right|}(r,z) e^{i\left|n\right|H_{\left|n\right|}(z)\theta} \nonumber\\
A_{\left|n\right|\theta}(r,z) &=& -\frac{\left|n\right| H_{\left|n\right|}(z)}{e_{|n|}^{eff}(z)} A_{\left|n\right|}(r,z).
\end{eqnarray}
Specifically, we can choose to define the function $H_{\left|n\right|}(z)$ as 
\begin{equation*}\label{eq:Hn1}
H_{\left|n\right|}(z) = \left \lbrace
\begin{array}{rl}
0 & \text{if } m\Delta - \left|n\right|l_p \leq z \leq m\Delta + \left|n\right|l_p \\
+1 & \text{if } 2m\Delta + \left|n\right|l_p < z < (2m+1)\Delta - \left|n\right|l_p \\
-1 & \text{if } (2m+1)\Delta - \left|n\right|l_p < z < (2m+2)\Delta + \left|n\right|l_p,
\end{array}\right.
\end{equation*}
where $m \in \mathbb{Z}$ and $\Delta$ is some scale characterising the length of a section of $\pm \left|n\right|$ 
string (which we expect to satisfy $\Delta \geq r_s,r_v$). 
We see that $H_{\left|n\right|}(z)$ admits a representation in terms of a superposition of Heaviside step functions $\Theta(z)$, 
specified over appropriate ranges, such that
\begin{equation*}\label{eq:Hn2}
H_{\left|n\right|}(z) = \left \lbrace
\begin{array}{rl}
& \Theta \left(z - 2m\Delta - \left|n\right|l_p\right), \ 2m\Delta - \left|n\right|l_p \leq z < (2m+1)\Delta - \left|n\right| l_p  \\
& \Theta \left(-\left(z - (2m+1)\Delta - \left|n\right| l_p\right)\right), \ 2m\Delta + \left|n\right| l_p < z \leq (2m+1)\Delta + \left|n\right| l_p \\
- & \Theta \left(z - (2m+1)\Delta - \left|n\right| l_p\right), \ 2(m+1)\Delta - \left|n\right| l_p \leq z < 2(m+1)\Delta - \left|n\right| l_p \\
- & \Theta \left(-\left(z - 2(m+1)\Delta - \left|n\right|l_p\right)\right), \  (2m+1)\Delta + \left|n\right| l_p < z \leq 2(m+1)\Delta + \left|n\right| l_p.
\end{array}\right.
\end{equation*}
For future reference we also note that the square of $H_{\left|n\right|}(z)$ is 
\begin{equation*}\label{eq:Hn3}
H^2_{\left|n\right|}(z) = \left \lbrace
\begin{array}{rl}
0 & \text{if } m\Delta - \left|n\right|l_p \le z \le m\Delta + \left|n\right|l_p \\
1 & \text{if } m\Delta + \left|n\right|l_p < z < (m+1)\Delta - \left|n\right|l_p,
\end{array}\right.
\end{equation*}
or equivalently
\begin{equation*}\label{eq:Hn4}
H^2_{\left|n\right|}(z) = \left \lbrace
\begin{array}{rl}
& \text{if } \Theta \left(z - m\Delta + \left|n\right| l_p\right), \ m\Delta - \left|n\right| l_p \leq z < (m+1)\Delta - \left|n\right| l_p \\
& \text{if } \Theta \left(-z - m\Delta - \left|n\right| l_p\right), \ m\Delta + \left|n\right| l_p < z \leq (m+1)\Delta + \left|n\right| l_p.
\end{array}\right.
\end{equation*}
The first derivative of $H_{\left|n\right|}(z)$, the square of its first derivative and its second derivative are then
\begin{eqnarray}\label{eq:Hn5}
\frac{dH_{\left|n\right|}(z)}{dz} 
&= & \sum_{m=-\infty}^{\infty} \left[\delta\left(z-2m\Delta - \left|n\right|l_p\right)
+ \delta\left(z-2m\Delta + \left|n\right| l_p\right)\right] \\
&-& \sum_{m=-\infty}^{\infty} \left[\delta\left(z-(2m+1)\Delta - \left|n\right| l_p\right)
+ \delta\left(z-(2m+1)\Delta + \left|n\right| l_p\right)\right], \nonumber
\end{eqnarray}
\begin{eqnarray}\label{eq:Hn6}
\left(\frac{dH_{\left|n\right|}(z)}{dz}\right)^2 
&=& \sum_{m=-\infty}^{\infty}\left[\delta^2\left(z-m\Delta - \left|n\right|l_p\right) + \delta^2\left(z-m\Delta + \left|n\right| l_p\right)\right],
\end{eqnarray}
\begin{eqnarray}\label{eq:Hn6}
\frac{d^2H_{\left|n\right|}(z)}{dz^2} 
&= & \sum_{m=-\infty}^{\infty} \left[\delta'\left(z-2m\Delta - \left|n\right|l_p\right)
+ \delta'\left(z-2m\Delta + \left|n\right| l_p\right)\right] \\
&-& \sum_{m=-\infty}^{\infty} \left[\delta'\left(z-(2m+1)\Delta - \left|n\right| l_p\right)
+ \delta'\left(z-(2m+1)\Delta + \left|n\right| l_p\right)\right], \nonumber
\end{eqnarray}
where the prime represents differentiation with respect to $z$. 
Although the mathematical definition of these functions is complicated, the functions themselves are easy to visualise 
and a plot of $H_{\left|n\right|}(z)$ is presented in Figure 1.
\begin{figure}[htp]
 \begin{center}
  \includegraphics[width=0.6\textwidth]{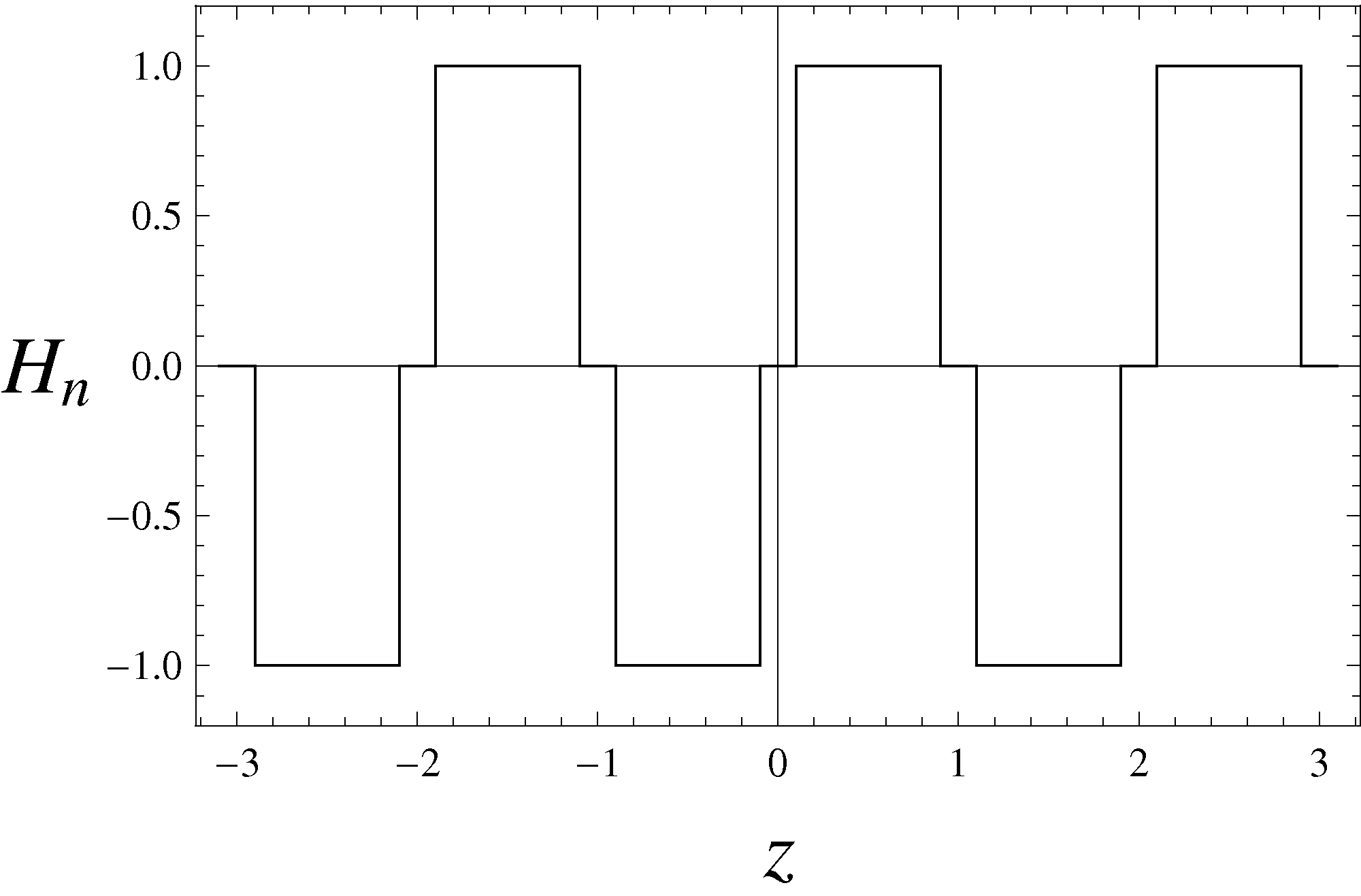} 
  \caption{$H_{\left|n\right|}(z)$ in the range $-3\Delta \leq z \leq 3\Delta$, with $\Delta=1$ and $|n|l_p=0.1$.}
 \end{center}
\end{figure}
Note that the Planck-sized regions where $H_{\left|n\right|}(z)=0$ do not represent genuine discontinuities 
in the phase of the complex field, or in the rotational direction of the gauge current. 
Rather they represent regions where both phase $(\theta)$ and the winding number $(\left|n\right|)$ 
(or equivalently the magnetic flux quantum number) are completely undefined, i.e. we have that,
\begin{eqnarray} \label{eq:undefined}
\theta = |n| = A_{|n|}(r,z) = 0, \ \ \ \forall r \leq |n|l_p, \ \ \ z \in [m\Delta \pm |n|l_p].  
\end{eqnarray}
It is not possible (or meaningful) to localise the string core on length scales $\Delta r < \mathcal{O}(l_p)$ or $\Delta z < \mathcal{O}(l_p)$. 
Furthermore, we may assume that only a change in topological number of $\pm 1$ will occur (on average) within any Planck-sized region, 
so that it is not possible to define continuous changes in $\left|n\right|$ over scales of $\Delta r < l_p$ or $\Delta z < l_p$, 
though we contend that this does not necessarily mean that discrete jumps in the topological winding number cannot take place over Planck-sized distances 
\footnote{Topological considerations prevent an $\left|n_1\right|$ vortex from morphing 
continuously into an $\left|n_2\right|$ vortex ($\left|n_1\right| \neq \left|n_2\right|$) over length scales $\Delta z >> l_p$ as there exists no 
homomorphism which smoothly maps one state to the other. Discontinuities are also usually considered unphysical. 
However on distances $\Delta z \sim l_p$ there exist no smooth maps at all. We argue that there is nothing, in principle, to prevent the 
topological winding number from changing discontinuously as long as it does so in a region where the discontinuity of space is also 
manifest i.e. over $\Delta z \sim \mathcal{O}(l_p)$.}.
The change from $+\left|n\right|$ to $-\left|n\right|$ winding states must therefore take place over a distance $\Delta z \approx 2\left|n\right| l_p$. 
The winding number $\left|n\right|$ is still undefined over the entire range $\Delta z \approx 2\left|n\right|l_p$, 
but a change in winding number of $\pm 2\left|n\right|$ cannot take place over a smaller distance. 
It is not, however, meaningful to ascribe a definite change in topological winding number of $\pm 1$ to a specific Planck length within $\Delta z$.
\\
\indent
It therefore seems natural to assume that $\left|n\right|$ may change (on average) by at most $\pm 1$ over a single Planck length $l_p$ 
and that the winding number is genuinely undefined at $r=0$ (where $\theta$ is also undefined) so that it may be set to zero without loss of generality. 
As we traverse the fundamental string core, the absolute value of the winding number changes from $\left|n\right|$ 
down to zero and then back up to $\left|n\right|$ which requires a distance of $\Delta r = 2\left|n\right|l_p$, so that
\begin{eqnarray} \label{eq:core_min}
r_s,r_v \geq \left|n\right| l_p
\end{eqnarray} 
The same bound can also be obtained in a different way: 
If we assume that is not meaningful for the phase of a complex field $\theta$ to vary continuously over Planckian distances, 
we may estimate the maximum rate of change with respect to the physical angular coordinate $\vartheta$ 
at the string core boundary $r_i$, $i \in \left\{s,v\right\}$, 
via $\left|\frac{d\theta}{d\vartheta}\right|_{r=r_i} \approx \frac{\Delta \theta}{\Delta \vartheta} \leq \frac{2\pi r_i}{l_p}$. 
But when $\Delta \vartheta = 2\pi$, we have $\Delta \theta = 2\pi \left|n\right|$ which recovers the relation above (\ref{eq:core_min}). 
This is the same as saying that, if the natural unit of phase is one radian, a phase change of $\Delta \theta = 1 \ rad$ 
cannot take place over a distance $\Delta s = r\Delta \theta < l_p$ on the circumference of a circle.
\footnote{
In the condensed matter analogue of the Abelian-Higgs model (c.f. \cite{Vilenkin+Shellard:book} and \cite{Zurek1}-\cite{Chuang_etal1}),
symmetry is dynamically broken when free electrons condense to form Cooper pairs 
resulting in a transition to a lower energy (superconducting) state with degenerate effective vacua in which magnetic flux lines 
(from any external magnetic field) are confined within localised flux tubes (see also \cite{Hook+Hall,superconductivity:book}). 
This implies that the Higgs field, which mediates spontaneous symmetry-breaking, may not be a fundamental field and may instead be 
interpreted as a phenomenological device which models the underlying (but as yet unknown) dynamical process. 
Formally, the order parameter of the Abelian-Higgs model (i.e. the phase $\theta$) is seen to be equivalent to the Cooper pair (Bogoliubov) 
wave-function of a superconducting fluid. Considering this in conjunction with the uncertainty principle 
then suggests that the minimum scale on which $\theta$ can be defined is $\sim l_p$. We then take this to imply that the minimum range over which graduated changes in the scalar field $\phi$ may occur is also $\sim l_p$. This applies equally to both the phase and magnitude of the complex field. Additionally the natural unit for the phase $\theta$ is one radian, which together with the considerations above, implies that a phase change of $\Delta \theta = \pm 1$ may not take place over a distance $\Delta r < l_p$ and hence gives rise to the condition (\ref{eq:core_min}).}
\subsection{Equations of motion for a pinched string} 

Turning now to the EOM for the non-cylindrical string, substituting the ansatz (\ref{eq:nscs_ansatz1}) 
into the appropriately modified versions of (\ref{eq:ELeqn1})-(\ref{eq:current2}) gives rise to the following scalar and vector equations:
\begin{eqnarray} \label{eq:nscs_scalar1}
0 &=& \frac{\partial^2 F_{|n|}}{\partial r^2} + \frac{1}{r}\frac{\partial F_{|n|}}{\partial r} + \frac{\left|n\right|^2H_{|n|}^2}{r^2}(A_{|n|}^2-1) 
- \frac{1}{(r_{s|n|}^{eff})^2}.\frac{1}{2} F_{|n|}(F_{|n|}^2-1) + \frac{\partial^2 F_{|n|}}{\partial z^2} \nonumber\\
&-& i\left|n\right| \theta \left\{F_{|n|} \frac{d^2 H_{|n|}}{d z^2} + 2\frac{\partial F_{|n|}}{\partial z}\frac{d H_{|n|}}{d z} \right\} 
- \left|n\right|^2 \theta^2 F_{|n|} \left(\frac{dH_{|n|}}{dz}\right)^2
\end{eqnarray}
\begin{eqnarray} \label{eq:nscs_vector1}
0 &=& H_{|n|}\left(\frac{\partial^2 A_{|n|}}{\partial r^2} - \frac{1}{r}\frac{\partial A_{|n|}}{\partial r} -\frac{|n|F_{|n|}^2}{(r_{v|n|}^{eff})^2}.(A_{|n|}-1)\right) \nonumber\\
&+& \frac{1}{r_{v|n|}^{eff}}\left\{r_{v|n|}^{eff}H_{|n|}\frac{\partial^2 A_{|n|}}{\partial z^2} + r_{v|n|}^{eff} \frac{d^2H_{|n|}}{dz^2}A_{|n|} + \frac{d^2r_{v|n|}^{eff}}{dz^2}H_{|n|}A_{|n|}\right\} \nonumber\\
&+& \frac{2}{r_{v|n|}^{eff}}\left\{r_{v|n|}^{eff} \frac{dH_{|n|}}{dz}A_{|n|} + \frac{dr_{v|n|}^{eff}}{dz}H_{|n|}A_{|n|} + r_{v|n|}^{eff}H_{|n|}\frac{\partial A_{|n|}}{\partial z} \right\}.
\end{eqnarray}
Considering (\ref{eq:nscs_scalar1}) first, we see that the imaginary part must be set equal to zero independently. 
However, the considerations above imply that the phase $\theta$ is effectively undefined not only at $r=0$ but for all $r \leq \left|n\right|l_p$ 
over the Planck-sized regions in which $H_{\left|n\right|}(z) = 0$, $z \in [m\Delta - \left|n\right|l_p, m\Delta + \left|n\right|l_p]$. 
We may therefore set $\theta = 0$ in this region without loss of generality, according to (\ref{eq:undefined}).
\\
\indent
In the regions where $H_{\left|n\right|}(z) = \pm 1$ ($z \notin [m\Delta - \left|n\right|l_p, m\Delta + \left|n\right|l_p]$), $\theta$
may be defined consistently but each term in the curly brackets must vanish independently and the imaginary component vanishes for all $z$. 
This argument is equivalent to multiplying the entire equation by $H_{\left|n\right|}(z)$ and setting either $H_{\left|n\right|}(z) = 0$ 
or the sum of terms which it multiplies equal to zero in alternate regions. 
Similar considerations hold for the term proportional to $\theta^2$, so that the final form of the scalar EOM simplifies to become
\begin{eqnarray} \label{eq:nscs_scalar2}
0 = \frac{\partial^2 F_{|n|}}{\partial r^2} + \frac{1}{r}\frac{\partial F_{|n|}}{\partial r} - \frac{\left|n\right|^2}{r^2}(A_{|n|}^2-1) 
+ \frac{1}{(r_{s|n|}^{eff})^2}.\frac{1}{2} F_{|n|}(F_{|n|}^2-1) + \frac{\partial^2 F_{|n|}}{\partial z^2} 
\end{eqnarray}
where we consider only the regions in which $H_{\left|n\right|}^2=1$. 
We may adopt the same strategy when dealing with the vector EOM, which shows that all terms in the curly brackets go to zero for all $z$ 
except the term proportional to $H_{|n|}\frac{\partial^2 A_{|n|}}{\partial z^2}$. This must be included in the final form of the EOM
\begin{eqnarray} \label{eq:nscs_vector2}
0 = \frac{\partial^2 A_{|n|}}{\partial r^2} - \frac{1}{r}\frac{\partial A_{|n|}}{\partial r} -\frac{|n|F_{|n|}^2}{(r_{v|n|}^{eff})^2}(A_{|n|}-1) + 
\frac{\partial^2 A_{|n|}}{\partial z^2} 
\end{eqnarray}
where again we need only consider regions in which $H_{\left|n\right|}^2=1$.
\\
\indent
This approach is equivalent to treating the Planck-scale regions as a ``black box" for which we have no effective field theory picture. 
Although we have no explicit expressions for the scalar and vector field functions $F_{|n|}$ and $A_{|n|}$, 
with which to calculate the tension of the string within the ranges $z \in [m\Delta \pm |n|l_p]$, 
we will later assume a tension of $\sim 2\pi \eta^2 |n|$ in these regions to ensure the continuity of $\mu_{|n|}(z)$. 
Although this is somewhat unsatisfactory, we will see that both the assumed tension for the Planck scale sections and 
the spatial localisation of the string core to scales $\sim |n|l_p$ have a natural explanation in the string theory picture, 
which we propose as a justification for the assumptions made here.
\\
\indent
Using our current approach it is impossible to obtain solutions for $F_{\left|n\right|}(r,z)$ and $A_{\left|n\right|}(r,z)$ 
in the regions for which $H_{\left|n\right|}^2=0$, and these remain untreated in our present analysis. 
Clearly if the solutions we obtain in the regions where $H_{\left|n\right|}^2=1$ are to be viewed as physical, 
the Planck-sized regions which connect sections of vortex/anti-vortex string must be dealt with in such a way as to ensure continuity, 
at least with respect to the string tension $\mu_{\left|n\right|}(z)$. 
This problem will be dealt with in the following section, in which $\mu_{\left|n\right|}(z)$ is calculated explicitly \footnote{Many thanks to V.M Red'kov 
for his insightful questions and comments regarding this point.}.
\\
\indent
Although we have yet to specify the precise form of the $z$-dependence of $F_{\left|n\right|}(r,z)$ 
and $A_{\left|n\right|}(r,z)$ (in the regions for which $H_{\left|n\right|}^2 \neq 0$) we may still use our physical intuition to impose appropriate boundary conditions. 
It is reasonable to assume that conditions analogous to those imposed on $f(r)$ and $\alpha(r)$ still hold for any
value of $z$, just as they did in the cylindrically symmetric case, so 
\begin{equation} \label{eq:F_bc2}
F_{\left|n\right|}(r,z) = \left \lbrace
\begin{array}{rl}
0 & \text{if } r=0 \hspace{1cm} \forall z \\ 
1 & \text{if } r \rightarrow \infty \hspace{0.74cm} \forall z 
\end{array}\right.
\end{equation}
and
\begin{equation}\label{eq:Alpha_bc2}
A_{\left|n\right|}(r,z) = \left \lbrace
\begin{array}{rl}
0 & \text{if } r=0 \hspace{1cm} \forall z \\ 
1 & \text{if } r \rightarrow \infty \hspace{0.74cm} \forall z.
\end{array}\right.
\end{equation}
The only problem that remains is how to specify ``large" and ``small" $r$ for fixed $z$. 
In the cylindrically symmetric case this was easily solved, since the length scales $r_s$ and $r_v$ determined the radii of the scalar and vector cores,
respectively, at every point along the string. 
In the non-cylindrically symmetric case we may expect the proposed substitutions (\ref{eq:subs1})-(\ref{eq:subs3}) to imply
the following boundary conditions on the related functions $F, A$:\\
\begin{equation}\label{eq:F_bc3}
F(R_{\left|n\right|s}^{eff}(z)) = \left \lbrace
\begin{array}{rl}
0 & \text{if } R_{\left|n\right|s}^{eff}(z) \rightarrow 0 \ \ \ \hspace{0.15cm} (r << r_{\left|n\right|s}^{eff}(z)) \\
1 & \text{if } R_{\left|n\right|s}^{eff}(z) \rightarrow \infty \ \ \  (r >> r_{\left|n\right|s}^{eff}(z))
\end{array}\right.
\end{equation}\\
and \\
\begin{equation}\label{eq:Alpha_bc3}
A(R_{\left|n\right|v}^{eff}(z)) = \left \lbrace
\begin{array}{rl}
0 & \text{if } R_{\left|n\right|v}^{eff}(z) \rightarrow 0 \ \ \  \hspace{0.15cm}(r << r_{\left|n\right|v}^{eff}(z)) \\
1 & \text{if } R_{\left|n\right|v}^{eff}(z) \rightarrow \infty \ \ \ (r >> r_{\left|n\right|v}^{eff}(z)).
\end{array}\right.
\end{equation}
We then need only specify the forms of $r_{\left|n\right|s}^{eff}(z)$ and $r_{\left|n\right|v}^{eff}(z)$ by introducing an appropriate 
additional ansatz, before verifying that they solve the EOM. To begin we note that
\begin{eqnarray}
\frac{\partial F_{|n|}}{\partial r} &=& \frac{1}{r_{s|n|}^{eff}}\frac{dF}{dR_{s|n|}^{eff}}, \nonumber\\
\frac{\partial^2 F_{|n|}}{\partial r^2} &=& \frac{1}{(r_{s|n|}^{eff})^2}\frac{d^2F}{d(R_{s|n|}^{eff})^2},
\end{eqnarray}
and
\begin{eqnarray}
\frac{\partial F_{|n|}}{\partial z} &=& -\frac{1}{r_{s|n|}^{eff}}R_{s|n|}^{eff} \frac{dr_{s|n|}^{eff}}{dz} 
\frac{dF}{dR_{s|n|}^{eff}}, \nonumber\\
\frac{\partial^2 F_{|n|}}{\partial z^2} &=& \frac{1}{r_{s|n|}^{eff}} R_{s|n|}^{eff}
\left[\frac{2}{r_{s|n|}^{eff}}\left(\frac{dr_{s|n|}^{eff}}{dz}\right)^2 - \frac{d^2r_{\left|n\right|s}^{eff}}{dz^2}\right]\frac{dF}{R_{\left|n\right|s}^{eff}} \nonumber\\
&+& \frac{1}{(r_{s|n|}^{eff})^2}(R_{s|n|}^{eff})^2 \left(\frac{dr_{s|n|}^{eff}}{dz}\right)^2 \frac{d^2F}{d(R_{s|n|}^{eff})^2},
\end{eqnarray}
and that similar relations exist between $A_{|n|}(r,v)$ and $A(R_{s|n|}^{eff})$. The scalar EOM then becomes
\begin{eqnarray} \label{eq:nscs_scalar2}
0 &=& \left[1 + (R_{s|n|}^{eff})^2 \left(\frac{dr_{s|n|}^{eff}}{dz}\right)^2\right] \frac{d^2F}{d(R_{s|n|}^{eff})^2} \nonumber\\
&+& \left[1 + (R_{s|n|}^{eff})^2\left\{2\left(\frac{dr_{s|n|}^{eff}}{dz}\right)^2 
- r_{s|n|}^{eff} \frac{d^2r_{s|n|}^{eff}}{dz^2}\right\}\right] \frac{1}{R_{s|n|}^{eff}} \frac{dF}{dR_{s|n|}^{eff}}
\nonumber\\
&+& \frac{|n|^2 F}{(R_{s|n|}^{eff})^2} - \frac{1}{2}F(F^2-1)\left(\frac{r_{s|n|}^{eff}}{r_s}\right).
\end{eqnarray}
By analogy with the cylindrically symmetric case, we may then write
\begin{eqnarray} \label{eq:nscs_scalar3}
0 &=& \left[1 + (R_{v|n|}^{eff})^2 \left(\frac{dr_{v|n|}^{eff}}{dz}\right)^2\right] \frac{d^2F}{d(R_{v|n|}^{eff})^2} \nonumber\\
&+& \left[1 + (R_{v|n|}^{eff})^2\left\{2\left(\frac{dr_{v|n|}^{eff}}{dz}\right)^2 
- r_{v|n|}^{eff} \frac{d^2r_{v|n|}^{eff}}{dz^2}\right\}\right] \frac{1}{R_{v|n|}^{eff}} \frac{dF}{dR_{v|n|}^{eff}}
\nonumber\\
&+& \frac{|n|^2 F}{(R_{v|n|}^{eff})^2} - \frac{1}{2}\beta_{|n|}^{eff}F(F^2-1)
\end{eqnarray}
where we have defined
\begin{eqnarray} \label{eq:beta_eff_1}
\beta_{\left|n\right|}^{eff}(z) = \left(\frac{r_{v|n|}^{eff}(z)}{r_{s|n|}^{eff}(z)}\right)^2.
\end{eqnarray}
Hence we see that in the case of the pinched string, the assumptions $F = F(R_{s|n|}^{eff})$ and $F = F(R_{v|n|}^{eff})$ are equivalent, 
in correspondence with the cylindrically symmetric case. 
The vector EOM becomes
\begin{eqnarray} \label{eq:nscs_vector2}
0 &=& \left[1 + (R_{v|n|}^{eff})^2 \left(\frac{dr_{v|n|}^{eff}}{dz}\right)^2\right] \frac{d^2A}{d(R_{v|n|}^{eff})^2} \nonumber\\
&+& \left[-1 + (R_{v|n|}^{eff})^2\left\{2\left(\frac{dr_{v|n|}^{eff}}{dz}\right)^2 
- r_{v|n|}^{eff} \frac{d^2r_{v|n|}^{eff}}{dz^2}\right\}\right] \frac{1}{R_{v|n|}^{eff}} \frac{dA}{dR_{v|n|}^{eff}}
\nonumber\\
&-& |n|F^2(A-1)\left(\frac{r_{v|n|}^{eff}}{r_{v}}\right)^2,
\end{eqnarray}
so that in the limit $r_{s|n|}^{eff}(z) \rightarrow r_s$, $r_{v|n|}^{eff}(z) \rightarrow r_{v}$ we recover the usual cylindrically symmetric equations (\ref{eq:scs_eom1})/(\ref{eq:scs_eom1_mod}) and (\ref{eq:scs_eom2}). 
\\
\indent
We now proceed by specifying the conditions we wish $r_{s|n|}^{eff}(z)$ and $r_{v|n|}^{eff}(z)$ to satisfy. 
Let us assume that each function $r_{\left|n\right|i}^{eff}(z)$, $i \in \left\{s,v\right\}$,
varies between two values $|n|l_p \leq r_{i|n|}^{eff}(z) \leq r_i$ (in accordance with conditions imposed along with the definitions (\ref{eq:subs0}) and (\ref{eq:subs1}), 
such that
\begin{eqnarray} \label{eq:rsneff1}
r_{i|n|}^{eff}\left(z=(m+1/2)\Delta \right) = r_i
\end{eqnarray}
and
\begin{eqnarray} \label{eq:rsneff2}
r_{i|n|}^{eff}(z) = \left|n\right| l_p, \ \ \ \forall z \in \left[m\Delta \pm |n|l_p\right].
\end{eqnarray}  
Additionally we impose the following constraint on the derivatives to ensure continuity at the points $z = m\Delta \pm |n|l_p$,
\begin{eqnarray} \label{eq:rsneff3}
\left|\frac{dr_{i|n|}^{eff}(z)}{dz}\right|_{z=m\Delta \pm |n|l_p} = 0,
\end{eqnarray} 
which together with (\ref{eq:rsneff2}) also implies
\begin{eqnarray} \label{eq:rsneff4}
\left|\frac{d^2r_{i|n|}^{eff}(z)}{dz^2}\right|_{z=m\Delta \pm |n|l_p} = 0.
\end{eqnarray} 
The most general form for the two functions $r_{\left|n\right|i}^{eff}(z)$ is 
\begin{eqnarray} \label{eq:r_in1}
r_{i|n|}^{eff}(z) = A_{i|n|} G_i\left(B_{i\left|n\right|}\left(z + C_{i\left|n\right|}\right)\right) + D_{i\left|n\right|}
\end{eqnarray}
where $\left\{G_i(z)\right\}$, $i \in \left\{s,v\right\}$ are periodic functions in $z$, valued between $0$ and $1$. 
The values of the constants $A_{i\left|n\right|}$, $B_{i\left|n\right|}$, $C_{i\left|n\right|}$ and $D_{i\left|n\right|}$ 
may then be uniquely determined by requiring $r_{i\left|n\right|}^{eff}(z)$ pass through the three 
points $\left(m\Delta + \left|n\right|l_p, \left|n\right|l_p\right)$, $\left((m+1/2)\Delta, 1\right)$ 
and $\left((m+1)\Delta - \left|n\right|l_p, \left|n\right|l_p\right) \ \forall m \in \mathbb{Z}$, 
together with the requirement that the first derivatives (with respect to $z$) 
are zero at the first and last of these points. 
Assuming the functions $G_i(z)$ both have natural period $\pi$, this gives
\begin{eqnarray} \label{eq:r_in2}
r_{i\left|n\right|}^{eff}(z) = \left(r_i - \left|n\right|l_p \right) G_i\left(\frac{z - m\Delta - \left|n\right|l_p}{\pi^{-1}\left(\Delta - 2\left|n\right|l_p \right)}\right) + \left|n\right|l_p.
\end{eqnarray}
The variation of the scalar and vector core profiles (assuming non-critical coupling $r_v \neq r_s)$ for a non-cylindrical ``pinched" 
string are illustrated in Figure 2 using $G_s(z) = G_v(z) = \sin^2(z)$ as an example ansatz.  
\begin{figure}[htp]
 \begin{center}
  \includegraphics[width=0.6\textwidth]{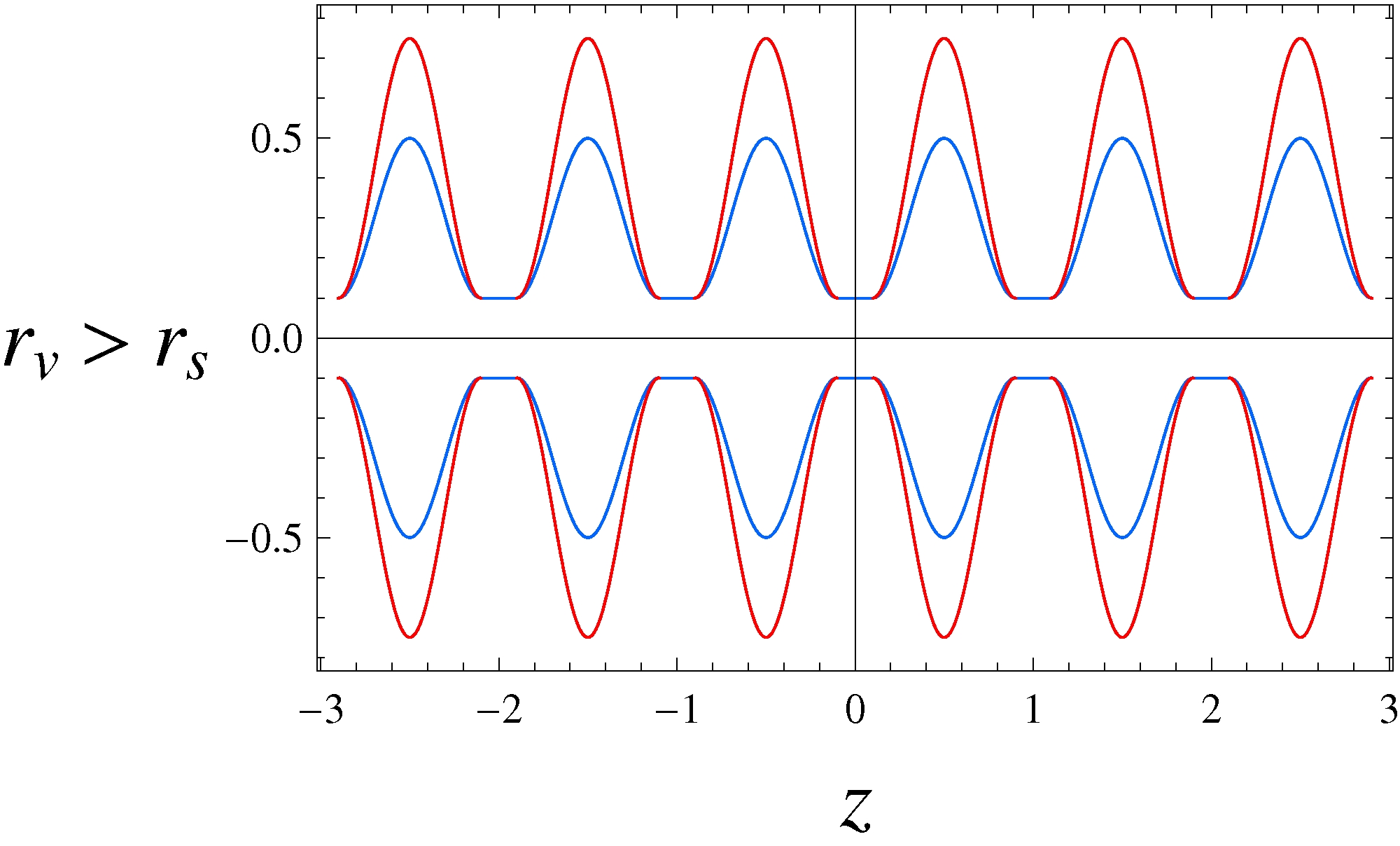} 
  \caption{Profile of the pinched string solution in the range $-3\Delta \leq z \leq 3\Delta$, with $\Delta= 1$,
$|n|l_p = 0.1$, $r_s = 0.5$ (blue curve) and $r_v = 0.75$ (red curve).}
 \end{center}
\end{figure}
For later convenience we note that for large $r_i$, satisfying $r_i >> \left|n\right|l_p$, we can make the approximation
\begin{eqnarray} \label{eq:r_in3}
r_{i}^{eff}(z) \approx r_i G_i\left(\frac{\pi z}{\Delta}\right) 
\end{eqnarray}
so that $r_{i\left|n\right|}^{eff}(z) \rightarrow r_{i}^{eff}(z)$ which becomes effectively independent of $\left|n\right|$ (except for the usual dependence on $\sqrt{|n|}$ of the vector core radius). This implies
\begin{eqnarray} \label{eq:r_in5}
\left|\frac{dG_i\left(\frac{\pi z}{\Delta}\right)}{dz}\right|_{z=m\Delta} = \left|\frac{dG_i\left(\frac{\pi z}{\Delta}\right)}{dz}\right|_{Z=(m+1/2)\Delta} = 0,
\end{eqnarray}
together with
\begin{eqnarray} \label{eq:r_in6}
\left|\frac{d^2G_i\left(\frac{\pi z}{\Delta}\right)}{dz^2}\right|_{z=m\Delta} > 0, \ \ \ \left|\frac{d^2G_i\left(\frac{\pi z}{\Delta}\right)}{dz^2}\right|_{z=(m+1/2)\Delta} < 0.
\end{eqnarray}  
In this limit, we may approximate $H_{\left|n\right|}(z)$ by the function $H(z)$, which is independent of $\left|n\right|$, and defined via
\begin{equation*}\label{eq:H}
H(z) = \left \lbrace
\begin{array}{rl}
& 2\Theta \left(z - 2m\Delta\right)-1, \ (2m-1)\Delta < z < (2m+1)\Delta  \\
& -2\Theta \left(z - (2m+1)\Delta \right) + 1, \ 2m\Delta < z < (2m+2)\Delta, \\
\end{array}\right.
\end{equation*}
whose square is given by
\begin{eqnarray}
H^2(z) = 1, \ \ \ \forall z
\end{eqnarray}
For future reference we note that
\begin{eqnarray}
\frac{dH}{dz} = \sum_{m=-\infty}^{\infty}\left[\delta(z-2m\Delta) - \delta(z-(2+1)m\Delta)\right],
\end{eqnarray}
\begin{eqnarray}
\frac{d^2 H}{d z^2} = \sum_{m=-\infty}^{\infty}\left[\delta'(z-2m\Delta) - \delta'(z-(2+1)m\Delta)\right],
\end{eqnarray}
and
\begin{eqnarray}
\left(\frac{dH}{dz}\right)^2 = \sum_{m=-\infty}^{\infty}\delta^2(z-m\Delta).
\end{eqnarray}
Additionally, note that in the limit $r_i >> |n|l_p$, we have $\beta_{\left|n\right|}^{eff}(z) \rightarrow \beta^{eff}(z)$ where
\begin{eqnarray} \label{eq:betaB}
\beta^{eff}(z) = \left(\frac{r_{v}^{eff}(z)}{r_{s}^{eff}(z)}\right)^2 
= \beta \left(\frac{G_v(z)}{G_s(z)}\right)^2.
\end{eqnarray}
If the $z$-dependence of $r_s$ and $r_v$ has the same functional form, i.e. if $G_s(z) = G_v(z) = G(z)$ so that
\begin{eqnarray} \label{eq:r_in4}
r_{i}^{eff}(z) \approx r_i G\left(\frac{\pi z}{\Delta}\right) 
\end{eqnarray} 
then the parameter $\beta^{eff}(z) \rightarrow \beta$ and becomes effectively independent of both $\left|n\right|$ and $z$. 
At critical coupling, therefore, it too is equal to unity. 
\\
\indent
Finally defining the dimensionless variable $Z$ through
\begin{equation}
Z \equiv \frac{\pi z}{\Delta}
\end{equation}
the functions $H(z)$, $r_{i}^{eff}(z) \approx r_i G_i(\pi z/\Delta)$ and $\beta^{eff}(z)$ may be rewritten as $H(Z)$, $r_{i}^{eff}(Z) \approx r_i G_i(Z)$ 
and $\beta^{eff}(Z)$ (this notation will be useful later).
\subsection{Solutions to the pinched string field equations}
We are now in a position to consider the solutions of (\ref{eq:nscs_scalar2})/(\ref{eq:nscs_scalar3}) and (\ref{eq:nscs_vector2}): 
If the terms in the square brackets in each equation satisfy
\begin{eqnarray} \label{eq:constraint1}
(R_{i|n|}^{eff})^2 \left(\frac{dr_{i|n|}^{eff}}{dz}\right)^2 \leq \mathcal{O}(1)
\end{eqnarray}
and 
\begin{eqnarray} \label{eq:constraint2}
(R_{i|n|}^{eff})^2\left\{2\left(\frac{dr_{i|n|}^{eff}}{dz}\right)^2 
- r_{i|n|}^{eff} \frac{d^2r_{i|n|}^{eff}}{dz^2}\right\} \leq \mathcal{O}(1)
\end{eqnarray}
within the region of the effective string core (i.e. $R_{i|n|}^{eff} \leq 1$, or equivalently $r \leq r_{i|n|}^{eff}$), 
then the EOM take functional forms which are identical (to within an order of magnitude in the coefficients of the derivative terms) 
to those for the cylindrically symmetric ansatz, under the correspondence (\ref{eq:subs1})-(\ref{eq:subs3}). 
They will therefore admit functionally equivalent solutions under the same correspondence in both the small $r$ and asymptotic limits, 
which may now be defined (for a given value of $z$) with respect to $r_{s|n|}^{eff}$ for the scalar EOM and $r_{v|n|}^{eff}$ for the vector EOM.
\\
\indent
Let us first consider the constraint (\ref{eq:constraint1}) which, using the definitions above, is approximately equivalent to
\begin{eqnarray}
R_{i|n|}^{eff} . \frac{r_i^2}{\Delta^2}(G_{i}')^2 \leq 1
\end{eqnarray} 
where $R_i^{eff} = r/r_i^{eff}$ and a prime denotes differentiation 
with respect to $Z=\pi z/\Delta$. 
Let us further assume that $G_{i}' \leq \mathcal{O}(1) \ \forall Z$ 
\footnote{In the absence of a specific ansatz for $G_i(z)$, 
this assumption may be questioned. Realistically it is necessary to input an initial, smoothly varying, function $G_i(z)$ 
and to subsequently calculate the initial forces acting on neighbouring string sections due to the interaction of the opposing magnetic fields. 
Ideally a dynamic theory will be developed which describes the evolution of the string core ``pinching" under the action of such forces. 
We may expect the pinching of the string to to become increasingly localised within the central area of the neighbouring $\pm\left|n\right|$ regions, 
resulting in ever-increasing localisation of the string energy density. Ultimately this would result in the formation of highly localised ``beads", 
and of neighbouring bead-antibead pairs which repel one another. 
Furthermore  each ``pinch" in the physical radius of the string gives rise to two peaks in the effective tension (i.e. two beads or antibeads) 
associated with the points of maximum absolute gradient $(dr_i^{eff}/dz)^2$. 
Increasing localisation of the pinching therefore corresponds to the coalescing of neighbouring $b-b$ and $\overline{b}-\overline{b}$ 
pairs as well as the repulsion of neighbouring $b$s and $\overline{b}$s. 
This could be the analogue of the creation of step-like windings in the formation of cosmic necklaces from smoothly wound strings, 
which gives rise to similar phenomena \cite{LakeThomasWard:JHEP2009}. If so, a future analysis of the interaction between neighbouring pinches 
may provide insights into the binding energy between neighbouring bead/antibead pairs in the wound string picture.}
giving
\begin{equation}\label{eq:oom_arg4}
\Delta \geq R_{i|n|}^{eff} r_i.
\end{equation}
Since we are interested in the physics at scales $R^{eff}_{i|n|} \leq \mathcal{O}(1)$, this constraint is essentially equivalent to the stronger condition
\begin{equation}\label{eq:oom_arg5}
\Delta \geq r_i,
\end{equation}
which appears reasonable on cosmic scales. As long as this condition is satisfied, we may be certain that the terms in the 
first set of square brackets (in both scalar and vector EOM) are $\sim \mathcal{O}(1)$ for all $r$ within the effective scalar and vector cores.
In the limiting case $r_s \sim \Delta$, this statement still holds. Another way to think about this is to note that the 
former constraint (\ref{eq:oom_arg4}) is equivalent to
\begin{equation} \label{eq:oom_arg4n2}
r \leq \Delta \frac{r_{\left|n\right|i}^{eff}(z)}{r_i}.
\end{equation}  
Since $r_{\left|n\right|i}(z) \in \left[\left|n\right|l_p, r_i\right] \ \forall z$, this effectively reduces to the minimum/maximum conditions
\begin{equation} \label{eq:oom_arg4n3}
r \leq \frac{\Delta \left|n\right|l_p}{r_i}, \ \ \ z \in[m\Delta \pm \left|n\right|l_p],
\end{equation}
or
\begin{equation} \label{eq:oom_arg4n3a}
r \leq \Delta, \ \ \ z \rightarrow (m+1/2)\Delta.
\end{equation}
Hence the condition $\Delta \geq r_i$ implies that (\ref{eq:oom_arg4}) is fulfilled for all $r \leq \left|n\right|l_p$ 
even at the boundaries of the region $z = m\Delta \pm \left|n\right|l_p$ (that is, for all $r$ within the effective radius at any value of $z$).
\\
\indent
It is also possible to identify the distance between neighbouring pinches, $\Delta$, with twice the (average) 
distance between peaks in the periodic tension, i.e. twice the average distance between neighbouring beads or anti-beads. 
In this case, by analogy with wound string necklaces, it is possible to identify this with twice the correlation length of the string, $\xi$, 
so that the number of strings per Hubble volume is $\nu \propto \Delta^{-3} \propto \xi^{-3}$ and the average distance between 
neighbouring strings is of order $\Delta$ \cite{LakeThomasWard:JHEP2009}. It is therefore likely that condition (\ref{eq:oom_arg4n3a}) will always be fulfilled
\footnote{However, in principle, $\Delta \sim r_s,r_v$, may occur, in which case it is doubtful whether the correspondence $\Delta \sim \xi$ 
may be maintained except shortly after the epoch of string network formation. Nonetheless in a fully dynamical 
theory of pinched string formation/evolution an explicit model of $\Delta(t)$ could be obtained, 
which may also allow the correspondence $\Delta(t) \sim \xi(t)$ to hold for all $t$.}.
\\
\indent
A similar argument can be used to show that the terms in the second set of square brackets are also of order one (for the scalar EOM) 
or of order minus one (for the vector EOM) for all $z$. 
Imposing the bound (\ref{eq:constraint2}) for all values of $z$ requires 
\begin{eqnarray} 
r^2\left\{2(G'_i)^2 - G''_i\right\} \leq \Delta^2 G_i^2.
\end{eqnarray} 
Now, assuming $G''_i(z) \sim G'_i(z) \sim \mathcal{O}(1) \ \forall z$, we have
\begin{eqnarray} 
r \leq \Delta G_i(z)
\end{eqnarray} 
which, accounting for the two limiting values of $r_{\left|n\right|i}^{eff}$, again gives rise to the minimum/maximum 
constraints (\ref{eq:oom_arg4n3})/(\ref{eq:oom_arg4n3a}) above.
\\
\indent
These arguments imply that the rate of change of $F_{|n|}\left(r,z\right)$ (or $F(R^{eff}_{s|n|}(z))$) 
and $A_{|n|}\left(r,z\right)$ (or $A(R^{eff}_{v|n|}(z))$) with respect to $z$ is sufficiently small 
at any value of $r$ so as to make $z$-derivative terms in the Euler-Lagrange equations negligible - 
so long as we are restricted to the ranges $r \leq r^{eff}_{s|n|}$ and $r \leq r^{eff}_{v|n|}$, respectively, and so long as the condition $r_s,r_v \leq \Delta$ holds. 
In principle, it is possible that the physics of non-cylindrically symmetric strings differs substantially from the symmetric case outside these ranges,
but this appears unlikely because the energy density of the field configuration rapidly goes to zero for $r > r^{eff}_{v|n|}(z), \ \forall z$ 
as $F \approx A \approx 1$. Additionally, we can identify $\Delta$ with the correlation length $\xi$, making $\Delta >> r_s,r_v$ a 
natural cut-off for $r$ in accordance with (\ref{eq:oom_arg4n3a}).
\\
\indent
It is also reasonable to assume $r_i \leq \Delta$ if we regard the pinch as a ``kink" or ``twist" in the string (analogous with our common sense ideas about ordinary lengths of string). The condition is then equivalent to the statement that it is impossible to twist a length of rope through 360 degrees over a region 
less than the length of its diameter.
In the limit $r_s \leq \Delta$ therefore, for all $r \leq r^{eff}_{s|n|}(z)$ the scalar EOM reduces to either
\begin{equation} \label{eq:nscs_scalar4}
0 = \frac{d^2F}{d(R_{s|n|}^{eff})^2} + \frac{1}{R_{s|n|}^{eff}} \frac{dF}{dR_{s|n|}^{eff}} 
+ \frac{|n|^2 F}{(R_{s|n|}^{eff})^2} - \frac{1}{2}F(F^2-1)
\end{equation}
or
\begin{equation} \label{eq:nscs_scalar5}
0 = \frac{d^2F}{d(R_{v|n|}^{eff})^2} + \frac{1}{R_{v|n|}^{eff}} \frac{dF}{dR_{v|n|}^{eff}} 
+ \frac{|n|^2 F}{(R_{v|n|}^{eff})^2} - \frac{1}{2}\beta_{\left|n\right|}^{eff}F(F^2-1).
\end{equation}
In the limit $r_v \leq \Delta$, \footnote{This is automatically satisfied for a 
type II superconducting regime $r_v > r_s$.} for all $r \leq r^{eff}_{\left|n\right|v}(z), \ (\forall z)$ the vector EOM reduces to
\begin{eqnarray} \label{eq:nscs_vector3}
0 = \frac{d^2A}{d(R_{v|n|}^{eff})^2} - \frac{1}{R_{v|n|}^{eff}} \frac{dA}{dR_{v|n|}^{eff}}
- |n|F^2(A-1)\left(\frac{r_{v|n|}^{eff}}{r_{v}}\right)^2.
\end{eqnarray}
In the uncoupled regime (i.e. setting $A=0$), equation (\ref{eq:nscs_scalar4}) has the small $r$ solution 
\begin{eqnarray} \label{eq:result1}
F(R^{eff}_{s|n|}) \approx (R^{eff}_{s|n|})^{|n|}
\end{eqnarray}
for $r \leq r^{eff}_{s|n|}(z)$, which is in accordance with our expectations regarding the effective radius of the scalar core. 
Likewise equation (\ref{eq:nscs_scalar5}) has the small $r$ solution
\begin{eqnarray}
F(R^{eff}_{\left|n\right|v}) &\approx& (\sqrt{\beta_{|n|}^{eff}}R^{eff}_{v|n|})^{\left|n\right|},
\end{eqnarray}
which is equivalent. Similarly, in the uncoupled regime (i.e. setting $F=0$) equation (\ref{eq:nscs_vector3}) admits the approximate solution for $A$
\begin{eqnarray} \label{eq:result2}
A(R^{eff}_{v|n|}) \approx (R^{eff}_{v|n|})^{2}
\end{eqnarray}
for $r \leq r^{eff}_{v|n|}(z)$. 
Clearly the correct asymptotic solutions in both uncoupled and coupled regimes (together with the higher order corrections to the small $r$ solutions in the latter) will also be formally analogous to those given for the cylindrical string in the usual literature, under the correspondence (\ref{eq:subs1})-(\ref{eq:subs3}). We need not state them explicitly here. More important for our immediate task are the results (\ref{eq:result1}) and (\ref{eq:result2}), which we now use to to calculate the periodic string tension $\mu_{|n|}(z)$ for the pinched string ansatz.
\subsection{Calculation of the (periodic) string tension for non-cylindrically symmetric string $\mu_{|n|}(z)$} 

When calculating the approximate $z$-dependent tension $\mu_{|n|}(z)$ for the pinched string, we begin by considering the 
limit $r_i >> |n|l_p$, $i \in \left\{s,v\right\}$.
\footnote{In the calculations below, we use the approximations $\sqrt{\lambda}_{|n|}^{eff}(z) \rightarrow \sqrt{\lambda}^{eff}(z)$ 
and $e_{|n|}^{eff}(z) \rightarrow e^{eff}(z)$ so that the field couplings are independent of $|n|$. 
This is equivalent to assuming $r_{v|n|}^{eff}(z) \rightarrow r_v^{eff}(z)$ and $r_{s|n|}^{eff}(z) \rightarrow r_s^{eff}(z)$ for $r_v,r_s >> |n|l_p$.}
There exist analogues of all the terms which appear in the (constant) string tension $\mu_{|n|}$ of a cylindrically symmetric string, according to the 
identification
\begin{eqnarray}
f &\longleftrightarrow& F, \nonumber\\
\alpha &\longleftrightarrow& A, \nonumber\\
r_i &\longleftrightarrow& r_i^{eff}(z), \nonumber\\
\beta &\longleftrightarrow& \beta^{eff}(z)
\end{eqnarray} 
and $d \longleftrightarrow \partial$ where necessary. \footnote{We have also used the fact that $H_{|n|}^2(z) \rightarrow H^2(z) = 1, \ \forall z$}
The $D_r$ and $D_{\theta}$ terms of the gradient energy give contributions of the form
\begin{eqnarray}
\int_{0}^{2\pi}d\theta \int_{0}^{r_s^{eff}(z)}D_r \phi \overline{D}^r \overline{\phi} r dr &+& \int_{0}^{2\pi}d\theta \int_{0}^{r_s^{eff}(z)}D_{\theta} \phi \overline{D}^{\theta} \overline{\phi} r dr 
\nonumber\\
&=& 2\pi \eta^2 \int_{0}^{r_s^{eff}(z)} \left(\frac{\partial F}{\partial r}\right)^2 rdr + 2\pi \eta^2 \int_{0}^{r_s^{eff}(z)} \frac{|n|^2 F^2}{r^2}|A^2-1|r dr 
\nonumber\\
&\sim& 2\pi \eta^2 |n|
\end{eqnarray}
In this case (again at leading order), exchanging $r_s \longleftrightarrow r_s^{eff}(z)$ makes no difference to the result of the 
calculation since all factors of either $r_s$ or $r_s^{eff}(z)$ exactly cancel in the final step. 
In the range $r_s^{eff}(z) \leq r \leq r_v^{eff}(z)$ the interchange of variables also makes no difference to the form of the 
contribution from the angular term, which is given by
\begin{eqnarray}
\int_{0}^{2\pi}d\theta \int_{r_s^{eff}(z)}^{r_v^{eff}(z)} \eta^2 \frac{|n|^2 F^2}{r^2}|A^2-1|r dr 
\sim 2\pi \eta^2 |n|^2 \ln\left(\sqrt{\beta^{eff}(z)}\right).
\end{eqnarray}
However there now exists a new $D_z$ term in the derivative, which leads to
\begin{eqnarray}
\int_{0}^{2\pi}d\theta \int_{0}^{r_s^{eff}(z)}\left|D_z \phi D^z \overline{\phi}\right| r dr 
&=& 2\pi \int_{0}^{r_s^{eff}(z)} \eta^2 \left(\frac{\partial F}{\partial z}\right)^2 r dr \nonumber \\
&\sim& \pi \eta^2 |n|^2 \frac{1}{1+|n|} \left(\frac{dr_s^{eff}}{dz}\right)^2 \nonumber \\
&\sim& \pi \eta^2 |n|  \times \frac{r_s^2}{\Delta^2} (G_s')^2, \ \ \ |n| \geq 1.
\end{eqnarray}
Likewise, the contribution from the potential term is formally equivalent under the identification $r_s \longleftrightarrow r_s^{eff}(z)$, which gives 
\begin{eqnarray}
\int_{0}^{2\pi}d\theta \int_{0}^{r_s^{eff}(z)} V(|\phi|)r dr &=& 2\pi \eta^2 \times \frac{1}{2(r_s^{eff})^2} \int_{0}^{r_s^{eff}(z)} F(F^2-1)^2 r dr \nonumber\\
&\sim& \frac{\pi \eta^2}{|n|+2}.
\end{eqnarray}
Turning our attention to the gauge field, the contribution from the $z$-component of the magnetic flux $B_z$ is analogous to that for the 
cylindrically symmetric case, so that
\begin{eqnarray}
\int_{0}^{2\pi}d\theta \int_{0}^{r_v^{eff}(z)} \frac{1}{2} B_zB^z r dr &=& \pi \times \frac{|n|^2}{2(e^{eff}(z))^2} \int_{0}^{r_v^{eff}(z)} \frac{1}{r^2} \left(\frac{\partial A}{\partial r}\right)^2 r dr \nonumber\\
&\sim& 2\pi \eta^2 |n|
\end{eqnarray} 
The only contribution left to calculate comes from the new radial $\vec{B}$-field component, $B_r = F_{z\theta}$. 
Our new gauge field ansatz is given by
\begin{eqnarray}
A_{|n|\theta} &=& -\frac{|n|H(z)}{e^{eff}(z)}A(r,z) \nonumber\\
&=& -\sqrt{2}\eta \sqrt{|n|} r^{eff}_v(z) H(z) A(r,z),
\end{eqnarray}
so 
\begin{eqnarray}
B_r = F_{z\theta} &=& \partial_z A_{\theta} \nonumber\\
&=&  -\sqrt{2}\eta \sqrt{|n|} \frac{\partial}{\partial z}(r^{eff}_v(z) H A) \nonumber\\
&=&  -\sqrt{2}\eta \sqrt{|n|} \left\{r^{eff}_v(z) \frac{\partial A}{\partial z}H + r^{eff}_v(z) A \frac{d H}{d z} + \frac{dr^{eff}_v}{dz}AH\right\}.
\end{eqnarray}
In the limit $r_v,r_s >> |n|l_p$, we have that $\frac{d H}{d z} = 0$ for all $z$ where $r^{eff}_v(z) \neq 0$ and $A \neq 0$ (i.e. for $z \neq m\Delta$), 
and $r^{eff}_v(z) = A = 0$ for all $z$ where $\frac{d H}{d z} \neq 0$ (i.e. at $z = m\Delta$).
This implies that the second term inside the brackets is zero for all $z$.
\footnote{Similar considerations hold true even if we neglect to take the limit $r_i >> |n|l_p$ and consider the Planck-sized regions explicitly.} 
The second term inside the curly brackets vanishes leaving
\begin{eqnarray}
B_r = -\sqrt{2}\eta \sqrt{|n|} \left\{r^{eff}_v(z) \frac{\partial A}{\partial z}H + \frac{dr^{eff}_v}{dz}AH\right\}
\end{eqnarray}
and
\begin{eqnarray} \label{eq:B_r_sqr}
\frac{1}{2}F_{z\theta}F^{z\theta} = \frac{|n|}{r^2} \left\{r^{eff}_v(z))^2 \left(\frac{\partial A}{\partial z}\right)^2 + 2r^{eff}_v(z)\frac{dr^{eff}_v(z)}{dz}A\frac{\partial A}{\partial z} + A^2\left(\frac{dr^{eff}_v(z)}{dz}\right)^2\right\}.
\end{eqnarray}
The first term inside the brackets of (\ref{eq:B_r_sqr}) yields
\begin{eqnarray}
2\pi \int_{0}^{r_v^{eff}(z)} \frac{|n|}{r^2} (r_v^{eff})^2 \left(\frac{\partial A}{\partial z}\right)^2 rdr 
&\sim& 8\pi \eta^2 |n| \frac{1}{(r_v^{eff}(z))^4} \left(\frac{dr^{eff}_v}{dz}\right)^2 \int_{0}^{r^{eff}_v} r^3 dr \nonumber\\
&\sim& 2\pi \eta^2 |n| \left(\frac{dr^{eff}_v}{dz}\right)^2 \nonumber\\
&\sim& 2\pi \eta^2 |n| \times \frac{r_v^2}{\Delta^2} (G'_v)^2
\end{eqnarray}
but this is exactly cancelled by the contribution of the second term
\begin{eqnarray}
2\pi \int_{0}^{r_v^{eff}(z)} \frac{|n|}{r^2} A \frac{\partial A}{\partial z} rdr 
&\sim& 2\pi \eta^2 |n| \times 2r_v^{eff}(z) \frac{dr^{eff}_v}{dz} \times -\frac{2}{r_v^{eff}(z)^5} \frac{dr^{eff}_v}{dz} \int_{0}^{r^{eff}_v} r^3 dr \nonumber\\
&\sim& -2\pi \eta^2 |n| \times \frac{r_v^2}{\Delta^2} (G'_v)^2,
\end{eqnarray}
so the only remaining contribution is the third term:
\begin{eqnarray}
2\pi \int_{0}^{r_v^{eff}(z)} \frac{|n|}{r^2} A^2 \left(\frac{dr_v^{eff}(z)}{d z}\right)^2 rdr 
&\sim& 2\pi \eta^2 |n| . \frac{1}{(r_v^{eff}(z))^4} \left(\frac{dr^{eff}_v}{dz}\right)^2 \int_{0}^{r_v^{eff}(z)} r^3 dr \nonumber\\
&\sim& \frac{\pi}{2} \eta^2 |n| \times \frac{r_v^2}{\Delta^2} (G'_v)^2.
\end{eqnarray}
The final approximate expression for the total string tension is then
\begin{eqnarray}
\mu_{|n|}(z) &\approx& 4\pi \eta^2 |n| + \pi \eta^2 |n|^2 \ln\left(\beta^{eff}(z)\right) + \pi \eta^2 |n| \times \frac{r_s^2}{\Delta^2} (G_s')^2 \nonumber\\
&+& \frac{\pi \eta^2}{|n|+2} + \frac{\pi}{2} \eta^2 |n| \times \frac{r_v^2}{\Delta^2} (G_v')^2.
\end{eqnarray}
Setting $G_s = G_v = G$ and $r_v = r_s = r_c$ (critical coupling) we then see that the large $|n|$ limit yields
\begin{eqnarray} \label{eq:finally}
\mu_{|n|}(z) \sim 4\pi \eta^2 |n| + \frac{3 \pi}{2} \eta^2 |n| \times \frac{r_c^2}{\Delta^2} (G')^2. 
\end{eqnarray}
We have assumed that the function $G(\ldots)$ has period $\pi$, so that $G(\pi z/\Delta)$ has period $\Delta$. 
This implies that $G'(\pi z/\Delta)$, and hence $G'(\pi z/\Delta)^2$, have period $\Delta/2$ so we may express the latter in the 
form $(G')^2 = (G')^2(2\pi z/\Delta)$. 
If we also assume that $G' \sim (G')^2 \sim \mathcal{O}(1) \ \forall z$, a simple and natural ansatz for $(G')^2$ is
\begin{eqnarray} \label{eq:G_ansatz}
(G')^2(z) = \frac{1}{2} \left[\sin^2\left(\frac{2\pi z}{\Delta}\right) + \sin^4\left(\frac{2\pi z}{\Delta}\right)\right]
\end{eqnarray}
though, in principle, there are an infinite number of possible ansatz choices. 
To within an order of magnitude therefore, our final expression for the approximate $z$-dependent tension of a pinched string is 
\footnote{There is a discrepancy between the numeric factors multiplying the periodic and constant parts of the tension. Since the only physical restriction on $G'$ is that $(G(z)')^2 \in [0,\mathcal{O}(1)] \ \forall z$, we have some freedom to define it to absorb factors of $2 \pi$. As mentioned previously, the numerical factor in front of the constant term is also to some degree arbitrary and depends upon the definition of $r_v$. We therefore assume that both can be adjusted to give $\mu_{|n|}(z) \approx 2\pi|n| + 2\pi|n| \times \frac{r_c^2}{\Delta^2} \times \tilde{G}(z)$ 
where $\tilde{G}(z)$ is some function which varies exactly between $\tilde{G}(z) \in [0,1]$.}
\begin{eqnarray} \label{eq:finally}
\mu_{|n|}(z) &\approx& 2\pi \eta^2 |n| + 2\pi \eta^2 |n| \times \frac{r_c^2}{\Delta^2} \times \frac{1}{2}\left[\sin^2\left(\frac{2\pi z}{\Delta}\right) + \sin^4\left(\frac{2\pi z}{\Delta}\right)\right].
\end{eqnarray}
In the next section we look at $F/D$-strings wrapping cycles around the $S^3$ manifold at the tip of the KS throat. 
From the Lagrangian of the theory we then determine an approximate formula for the effective four-dimensional tension of the configuration. 
We find that, for an appropriate natural ansatz choice for the winding state, this is formally analogous to the result (\ref{eq:finally}) 
and this allows us to draw a correspondence between the parameters which define the 
Abelian-Higgs model and those which define the tip geometry of KS background. This ensures that we can connect the string theory
model to the field theory one.
\subsection{Relation of the pinched string tension to wound $F$/$D$-strings} 
For the sake of brevity we refer the interested reader to \cite{LakeThomasWard:JHEP2009, LakeThomasWard:JCAP2010} 
for more background on the string theory construction. We note that cosmic superstrings 
\cite{Copeland:2005cy, Polchinski:2007qc} have 
become an important area of cosmological research in recent years. A selection of papers relating to the
observational status of such objects is provided in \cite{Bevis:2006mj}-\cite{Firouzjahi:2010ji}.
Our  ansatz (below) describes a static string loop of radius $\rho$
with integer windings over all three angular directions of the $S^3$ at the tip of the warped deformed conifold
\begin{eqnarray}
X^{\mu} = (t, \rho \sin\sigma, \rho \cos\sigma, z_0, 0, 0, 0, 0, \psi(\sigma) = n_{\psi}\sigma, \theta(\sigma) = n_{\theta}\sigma, \phi(\sigma) = n_{\phi}\sigma).
\end{eqnarray}
We have chosen to label the loop radius $\rho$ instead of $r$, in order to avoid confusion with the $r$-coordinate used in the previous section. 
\\
\indent
For a string with no intrinsic world-sheet flux $F_{ab}$ (i.e. $\Pi^2=0$, where $\Pi$ is the associated momentum) 
and adopting canonical coordinates so that the ansatz above describes non-geodesic 
windings, the total energy is given by the following potential
\begin{eqnarray} \label{eq:V1}
V = a_0 T_1 \int d\sigma \sqrt{a_0^2 \rho^2 + R^2(n_{\psi}^2 + \sin^2(n_{\psi}\sigma)(n_{\theta}^2 + \sin^4(n_{\theta}\sigma)n_{\phi}^2))}
\end{eqnarray}
where $T_1$ denotes the tension of either an $F$ or $D$-string (depending on the configuration under consideration). 
In \cite{LakeThomasWard:JHEP2009} the total string mass was separated into a constant piece - corresponding to the mass of the string sections connecting the ``beads" 
formed by extra-dimensional windings - and a piece due to the mass of the beads themselves. 
It is necessary to set $n_{\psi}=0$ when calculating the mass of an individual bead,
but to use $N = 2n_{\psi} > 0$ when calculating the number of beads in the loop. 
This avoids double counting the mass-contribution of the $\psi$-direction windings. 
We now perform the same procedure when calculating the effective four-dimensional tension.
\footnote{Although seemingly counter-intuitive this was a necessary step in the case of true necklace configuration, 
when the bead mass is almost totally localised in space. We therefore expect it to be necessary here even if the beads are less localised. 
Since we require $n_{\psi}>0$ in order for beads, or fluctuations, in the effective four-dimensional mass-density to occur, 
we must still set $n_{\psi} \sim 0$ when calculating the approximate mass contained in localised areas (see \cite{LakeThomasWard:JHEP2009} for further details).}
\\
\indent
We begin by setting $n_{\psi} \sim n_{\theta} \sim n_{\phi} \sim n_w$
\footnote{Here $n_w$ labels the number of windings, in order to avoid confusion with $n$, the topological winding number of the field theory vortex strings.} 
in the expression (\ref{eq:V1}) above, except in the constant piece (which is proportional to $n_{\psi}^2$), which again vanishes. 
Taking the limit $a_0^2 \rho^2 >> n_w^2 R^2$ we then expand our expression for $V$ (to first order) before 
dividing through by $a_0\rho$ to obtain 
\begin{equation} \label{eq:mu_sigma}
\mu(\sigma) \approx 2\pi a_0 T_1  + 2\pi a_0 T_1 \times \frac{1}{2}\frac{n_w^2 R^2}{a_0^2 \rho^2} \left[\sin^2(n_w\sigma) + \sin^4(n_w\sigma)\right].
\end{equation}
We may now make a change of variables via
\begin{equation}
a_0z = a_0 \rho \sigma
\end{equation}
and define a new variable
\begin{equation}
a_0d = \frac{2\pi a_0 \rho}{n_w}
\end{equation}
which represents the interbead distance (or equivalently the interwinding distance, i.e. the four-dimensional length over which a single winding is ``spread"). This gives
\begin{eqnarray} \label{eq:mu_z}
\mu(z) \sim 2\pi a_0 T_1  + 2\pi a_0 T_1 \times \frac{1}{2}\frac{R^2}{a_0^2d^2} \sin^2\left(\frac{2\pi z}{a_0d}\right),
\end{eqnarray}
which is formally equivalent to (\ref{eq:finally}) under the identification
\begin{eqnarray} \label{eq:correspondence1}
a_0 T_1 &\longleftrightarrow& \eta^2 |n|, \nonumber\\
R &\longleftrightarrow& r_c, \nonumber\\
a_0d &\longleftrightarrow& \Delta.
\end{eqnarray}
Physically what is happening (in the string picture) is the following: 
When $z = d/2$ the string is instantaneously wrapping its maximal effective radius $R^{eff}(z) =R$ in the $S^3$ (a great circle), 
whereas at $z=d$ the string is instantaneously wrapping the pole of the $S^3$ (a point) 
so that the effective radius of the winding is $R^{eff}(z)=0$ 
\footnote{However at \emph{both} these points, $\frac{dR^{eff}(z)}{dz} = 0$, so that the effective four-dimensional tension is simply $\mu(z) \approx 2\pi a_0 T_1$.}.
At all values of $z$ the string wraps some effective radius in the region $0 \leq R^{eff}(z) \leq R$, and the maximum increase in winding rate 
with respect to the $z$-coordinate, $\frac{dR^{eff}(z)}{dz} \sim \frac{dr_c^{eff}(z)}{dz}$, occurs at the points $z=d/4,3d/4$. 
From the wound string perspective, these are points at which the greatest length of string is hidden in the compact space (for a given interval $dz$), 
giving rise to maxima in the effective four-dimensional tension. Likewise the periodic part of the tension in the pinched string picture 
is proportional to $\left(\frac{dr_c^{eff}(z)}{dz}\right)^2$ and we make the following identification 
\footnote{There exist \emph{two} degenerate minima in the fundamental domain of the $\psi$-coordinate of the $S^3$ giving rise to two beads per winding. 
If one full winding in the string picture corresponds to one full pinch in the field picture, 
we would therefore expect to find two peaks in the periodic part of the tension $\mu_{|n|}(z)$ for every one peak in the physical radius of the string core.}
\begin{eqnarray}
R^{eff}(z) \longleftrightarrow r_c^{eff}(z).
\end{eqnarray}
which explains why we must set $r_v^{eff}(z) = r_s^{eff}(z) \ \forall z$ in the field theory picture 
to obtain (\ref{eq:correspondence1}) - in this picture there is only one string, which cannot give rise to two separate radii. 
This allows us to naturally interpret the $z$-dependence of the field couplings $\sqrt{\lambda}^{eff}(z)$ and $e^{eff}(z)$ via the relation
\begin{eqnarray}
R^2 \sim b_0 g_s M \alpha', \hspace{0.5cm} b_0 \sim \mathcal{O}(1)
\end{eqnarray}
The parameter $M$ is fixed and quantised, and $\sqrt{\alpha'}$ represents the fundamental length scale of the theory,
so that a string wrapping an effective radius in the $S^3$, $R^{eff}(z) \leq R$, experiences an effective coupling of approximately 
\begin{eqnarray}
g_s^{eff}(z) = (R^{eff}(z))^2/b_0M\alpha'.
\end{eqnarray}
This also allows us to establish a relation between $g_s^{eff}(z)$, $\sqrt{\lambda}^{eff}(z)$ and $e^{eff}(z)$, which will be investigated shortly.
\\
\indent
We can imagine a situation where the string completes one full winding in (say) the clockwise direction 
before reversing to wrap the $S^3$ anti-clockwise. The four-dimensional regions over which these windings take place may 
correspond to regions of $\pm |n|$ in the field theory picture. 
Although we still have no effective description for the Planck-sized regions of the pinched string in classical field theory, 
within the string picture we see that there is a relatively natural interpretation, with $|r_c^{eff}|_{min} = |n|l_p \equiv |R^{eff}|_{min}$ 
corresponding to the minimum width of the string due to quantum effects.
\\
\indent
Effectively when the string wraps a point (at the pole) of the $S^3$ at $z=d$, 
the description of the string as a one-dimensional object breaks down. 
It is therefore meaningless to consider the position of the string localised on the $S^3$ on scales smaller than the fundamental string width. 
Furthermore the four-dimensional effective tension of the string in this region is equal to the intrinsic warped tension $\tilde{T}_1 \sim a_0 T_1$, 
which, under the identification in (\ref{eq:correspondence1}), is equivalent to the tension of a non-cylindrical defect 
string in the region of the pinch, $\mu_{|n|}(z \approx m\Delta) \sim \eta^2 |n|$. 
This goes some way towards justifying our original assumption that $\mu_{|n|}(z) \sim \eta^2 |n|$ within the regions $z \in [m\Delta \pm |n|l_p]$.
\\
\indent
How tightly wound the string is may affect the four-dimensional length required for it to move from one winding orientation to another. 
If, for example, we identify the Planck length $l_p$ with the fundamental string length $l_s \sim \sqrt{\alpha'}$,
the fact that $|r_c^{eff}|_{min} \propto |n|$ (in addition to $l_p \equiv \sqrt{\alpha'}$) may be explained.
Intuitively we expect a more tightly wound string to give a larger value of $|n|$ in the field picture, 
so that the simple identification $n_w \longleftrightarrow |n|$ must be rejected. 
We can, however, arrive at a hypothetical correspondence between $|n|$ and the dynamical parameters which control winding formation in the 
string picture via the following argument: 
Recall that in the field theory picture we have assumed the sections of ``neutral" string, which connect neighbouring regions of $\pm |n|$ string, 
are of length $\sim 2|n|l_p$. 
We may now check this in the dual string picture by assuming that the four-dimensional length over which the string ``sits" at the pole of the $S^3$ is proportional to (twice) the tangent angle of incidence (i.e. a string which is more tightly ``wound" requires a greater distance over which to ``unwind", in direct proportion to its angle of incidence), giving
\begin{eqnarray}
\Delta z \sim 2|n|l_p \propto 2\frac{R}{a_0d}\sqrt{\alpha'}.
\end{eqnarray}
Identifying $l_p \sim \sqrt{\alpha'}$ we then have 
\begin{eqnarray}
|n| \propto \frac{R}{a_0d}.
\end{eqnarray}
However, this cannot be our final expression, since for $a_0d > R$, $|n| < 1$. 
Using the definition of $\omega_l$
\begin{eqnarray}
\omega_l \sim \frac{n_w R}{\sqrt{a_0^2 \rho^2 + n_w^2 R^2}} \sim \frac{R}{a_0d}\left(1 + \frac{R^2}{a_0^2d^2}\right)^{-\frac{1}{2}}
\end{eqnarray}
we see that $\omega_l \rightarrow 0$ as $a_0 d \rightarrow \infty$. We therefore propose the identification
\begin{eqnarray} \label{eq:n_correspondence}
|n| \sim \frac{R}{\omega_l a_0 d} \sim \frac{n_w R}{a_0\rho \omega_l} \sim \sqrt{1 + \frac{R^2}{a_0^2 d^2}} \sim \frac{1}{\sqrt{1-\omega_l^2}},
\end{eqnarray}
which implies $|n| \rightarrow \infty$ as $\omega_l \rightarrow 1$ and $|n| \rightarrow 1$ as $\omega_l \rightarrow 0$. 
Thus, in the limit that we obtain an unwound $F$-string, the original duality proposed by Nielsen and Olesen \cite{Nielsen_Olesen} is recovered
\footnote{For $|n|$ to be an integer greater than one (which corresponds to the limit $a_0d>>R$) we must ensure that $a_0d < R$, 
which is equivalent to $\Delta < r_c$ in the pinched string picture. Hence models in which $|n|>1$ are potentially problematic, since the assumptions made in order to simplify the pinched string EOM break down. This suggests that further analysis is needed, at least regarding the field-theoretic necklace model.}.
Topological winding numbers of opposite signs ($\pm n$) way be obtained by taking either physical winding numbers of opposite signs ($\pm n_w$) or
opposite signs in front of the square root.
\\
\indent
We now consider turning on world-volume flux. For an $F$-string, this amounts to turning on $D$-string charge, and therefore the resulting configuration
is a bound-state of $F$ and $D$-strings, known as a $(p,q)$ string. The tension of general $(p,q)$-string at the tip of this particular warped throat is 
\cite{Firouzjahi:2006vp}-\cite{Firouzjahi:2006xa},
\begin{eqnarray}
T_{(p,q)} \simeq \frac{1}{2\pi \alpha'}\sqrt{\left(\frac{q}{g_s}\right)^2 + \left(\frac{b_0M}{\pi}\right)^2 \sin^2\left(\frac{p\pi}{M}\right)}
\end{eqnarray}
so that in the limit $M>>1$, the approximate tension of the $F$-string is
\begin{eqnarray} \label{eq:F_string_tension}
T_{(1,0)} &\sim& \frac{1}{2\pi \alpha'}\left(\frac{b_0M}{\pi}\right)\sin\left(\frac{p\pi}{M}\right) \nonumber\\
&\sim& \frac{1}{2\pi \alpha'} \times \frac{b_0M}{\pi} \times \frac{\pi}{M}, \hspace{0.5cm} M>>1 \nonumber\\
&\sim& \alpha'^{-1},
\end{eqnarray}
and that of the $D$-string is
\begin{eqnarray} \label{eq:D_string_tension}
T_{(0,1)}  \sim  \frac{1}{2\pi \alpha'} \frac{1}{g_s} 
\sim \alpha'^{-1} g_s^{-1}
\end{eqnarray}
Dealing first with the $F$-string, we see that under the identification given in (\ref{eq:correspondence1}); 
the only consistent identification between the individual elements $\eta$ and $|n|$ is
\footnote{The alternative possibility is that it may not be possible to identify $\eta$ and $|n|$ directly with string theory parameters, but 
only through composite expressions of the form $\eta |n| \sim f(a_0,g_s,M)$ (e.g. $\eta^2 |n| \sim a_0 \alpha'^{-1}$ 
as suggested by the expression for the $F$-string tension above.)}
\begin{eqnarray} \label{eq:F_string_correspondence1}
\eta \sim a_0 \sqrt{\alpha'}^{-1}, \ \ \ |n| \sim \frac{1}{a_0}.
\end{eqnarray}
In \cite{LakeThomasWard:JHEP2009} it was necessary to identify the energy corresponding to the epoch of the $(p,q)$-string network 
formation ($\eta_s$) with the fundamental string energy scale (not the warped string energy scale) so that, $\eta_s \sim \alpha'^{-1/2}$ 
not $\eta_s \sim a_0\alpha'^{-1/2}$. Although this appears to contradict our identification (\ref{eq:F_string_correspondence1}) 
it is not immediately clear that this is the case since we have chosen to identify the fundamental string 
width $\delta \sim \eta_s^{-1}$ with the fundamental string scale (so that $\delta \sim \eta_s^{-1} \equiv l_s \sim \sqrt{\alpha'}$) 
which is in agreement with the previous results. 
The identification of $\eta$ with the warped string scale tells us that it is inequivalent to $\eta_s$. In the string picture the $F/D$-string network
forms at $t_s \sim \eta_s^{-1} \sim l_s$, but the windings only form some time later at $t_w > t_s$ where $t_w \sim a_0^{-1}l_s$ at which point
wound strings are dual to defect strings with $|n| \geq 1$. 
\\
\indent
The effective formation time may or may not be directly dependent on $a_0$ as this will be determined by the winding formation mechanism and the type of winding formed. For example, consider the expression for the the total energy of the wound-string loop for geodesic windings formed via the velocity correlations regime, 
$E = 2\pi T_1 \rho = 2\pi T_1 (\alpha t_i)$, where $T_1$ denotes the tension of either the $F$ or $D$-string (see \cite{LakeThomasWard:JCAP2010}). 
Using the arguments given above, geodesic windings in the string picture correspond to ``un-pinched" (i.e. cylindrically symmetric) strings in the dual field theory model, whose total energy we expect to be $E = 2\pi \eta^2 |n| \rho$ and whose (constant) tension is $\mu = 2\pi \eta^2 |n|$ 
(where $|n| \sim 1/\sqrt{1-\omega_l^2}$ as in (\ref{eq:n_correspondence})). This suggests the correspondence\footnote{Note that we must use the identification $\mu \sim E/(a_0\rho)$ in the wound-string case and $\mu \sim E/\rho$ in the field theory case.}
\begin{eqnarray} \label{eq:F_string_correspondence2}
\eta \sim \sqrt{\alpha'}^{-1}, \ \ \ |n| \sim \frac{1}{a_0}.
\end{eqnarray}
In (\ref{eq:F_string_correspondence1}) the symmetry-breaking energy scale is set by the warped string energy scale, 
whereas in (\ref{eq:F_string_correspondence2}) it is set by the fundamental string energy scale. It is unclear whether the string width should be set by the fundamental string length-scale or the warped string length-scale, 
and it is similarly unclear whether the energy associated with $F$/$D$-string network production should be the fundamental string or the warped string energy. 
Furthermore, it is unclear whether it is possible for the string width to be determined by the fundamental scale (i.e. $\sim \sqrt{\alpha'}$) 
while the energy associated with network formation (in the string picture)/symmetry-breaking (in the field theory picture) 
is determined by the warped string scale ($\sim a_0 \sqrt{\alpha'}^{-1}$), or vice-versa. Clearly therefore, more work is needed to establish the exact relation between the type of wound-string network formed, the winding formation mechanism (e.g. via a random walk or velocity correlations regime \cite{Avgoustidis:2005vm}) and the effective formation time. However, we note that the correspondence $|n| \sim 1/\sqrt{1-\omega_l^2} \sim 1/a_0$ in both (\ref{eq:F_string_correspondence1}) and (\ref{eq:F_string_correspondence2}) is encouraging, since in \cite{LakeThomasWard:JHEP2009} it was found that, for windings formed via a velocity correlations regime at least, $\omega_l \sim \sqrt{1-a_0^2}$, which is consistent with these results \cite{LakeThomasWard:JCAP2010}.
Considering the $D$-string (for non-geodesic windings) also leads to a unique correspondence between the individual field theory and string theory parameters:
\begin{eqnarray} \label{eq:D_string_correspondence2}
\eta \sim a_0\sqrt{\alpha'g_s}^{-1}, \ \ \ |n| \sim \frac{1}{a_0}
\end{eqnarray}
which preserves this key result. A relation analogous to (\ref{eq:F_string_correspondence2}) also holds for geodesic windings.
\\
\indent
Combining the expression $R^2 \sim g_s M \alpha'$ with 
$R^2 \sim r_s^2 \sim (\sqrt{\lambda} \eta)^{-2} \sim r_v^2 \sim |n|(e\eta)^{-2}$ and (\ref{eq:F_string_correspondence1}) gives
\begin{eqnarray} \label{eq:Izzy}
\lambda \sim e^2/|n| \sim \frac{1}{a_0^2 Mg_s} \sim \frac{\alpha'}{a_0^2 R^2}
\end{eqnarray}
and hence 
\begin{eqnarray}
\lambda^{eff}(z) \sim (e^{eff}(z))^2/|n| \sim \frac{1}{a_0^2 Mg_s^{eff}(z)} \sim \frac{\alpha'}{a_0^2 (R^{eff}(z))^2}.
\end{eqnarray}
Using the definition of the conifold deformation parameter
(and adjusting the units so that $\tilde{\epsilon}^{-4/3} = \epsilon^{4/3} \alpha'$), we may then write
\begin{eqnarray} \label{eq:final_correspondence}
\lambda \sim e^2/|n| \sim \frac{1}{a_0^2 Mg_s} \sim \frac{\alpha'}{a_0^2 R^2} \sim \epsilon^{4/3}.
\end{eqnarray}
On the left hand side of (\ref{eq:final_correspondence}) we have the Abelian-Higgs couplings, which together with the symmetry-breaking energy scale $\eta$, 
determine the effective masses of the particles according to that model (i.e. the bosons associated with scalar and vector fields). 
On the right hand side we have string theory parameters which determine the geometry according to the KS model 
(including both the large and compact dimensions) and which control the fundamental mass-scales of the particles associated with that theory 
(i.e. the masses associated with the excitations of $F$/$D$-strings). 
If we also include $\eta \sim a_0\sqrt{\alpha'}^{-1}$, the field theory parameters which set the mass scales for the particles at the current (post-symmetry breaking) epoch may be equated with the parameters which control inflation and the cosmological expansion in the string picture.
\\
\indent
At first glance this may seem strange. However, the decay of the inflaton gives rise to production of particles and defects
whose remnants or ``descendents" (following temperature changes and further symmetry-breaking phase transitions caused by the expansion and cooling of the universe) may be observed today. On reflection, the identification proposed above appears natural since we expect the parameters which control 
inflation to set the present day mass-scale of the universe.
\\
\indent
There are two important remarks to make about the above analysis.
The first regards the nature of the symmetry-breaking phase transition in the string picture. 
In the Abelian-Higgs model the phase transition, giving rise to string formation, is well-defined 
and it is the $U(1)$ symmetry of the vacuum that is broken. However
such symmetry-breaking should be dynamic, and moreover the coupling constants are tuned by hand.
According to our argument, we can equate the parameters determining the size and shape of the Mexican hat potential with their string analogues.
But how are we to interpret the symmetry-breaking on the string side?
We imagine a long, straight (i.e. unwound) $F/D$-string obeys $U(1)$ symmetry with respect to rotation around its central axis. 
When windings form this symmetry is broken and each point along the string adopts a ``phase" factor, determined by it's position in a $(U(1) \equiv S^1)$ 
sub-manifold of the full three-sphere
\footnote{Rotating the string around its fundamental axis will then no longer leave the configuration invariant. 
It is analogous to turning a screw, causing windings to change their position in the large dimensions, 
just as the threads of a screw move along its axis when the screw is turned.}.
In the string picture the dynamic nature of the symmetry-breaking process in therefore manifest, and helps us understand why we may write all 
the parameters which define the field-theoretic strings (including $|n|$) in terms of parameters controlling the $F$/$D$-string dynamics, 
and in particular those which control the process of winding formation (regardless of the exact model of winding formation we adopt).
\\
\indent
The second remark concerns cosmic necklaces, first proposed by Matsuda as possible DM candidates \cite{Matsuda:2005ez, Matsuda:2005fb}. 
The idea was that necklaces which had shrunk to their minimal size (determined by the fundamental string thickness) 
and which contained insufficient mass to undergo collapse, would only interact with other fields/matter gravitationally. 
The proposed correspondence between necklace configurations and pinched gauge strings now suggests that this is unlikely to occur, since 
the presence of a dual $A_{\theta}$ term implies that necklaces may also interact with the gauge field.
With this in mind, we propose the following definitions for the effective scalar and vector fields in the wound string model,
\begin{eqnarray}
\phi(r,\theta,z) &=& \phi(r,\theta,r_s^{eff}(z)) = \eta F\left(\frac{r}{r_s^{eff}(z)}\right) e^{\pm i |n| \theta} \nonumber\\
&\equiv& \frac{a_0}{\sqrt{\alpha'}} F\left(\frac{r}{R^{eff}(z)}\right) \exp\left(\pm i\left|\frac{n_w R}{\omega_l a_0 \rho}\right| \theta \right)
\end{eqnarray}
and 
\begin{eqnarray}
A_{\theta}(r,z) &=& A_{\theta}(r,r_s^{eff}(z)) = \eta r_v^{eff}(z) \sqrt{|n|} A\left(\frac{r}{r_v^{eff}(z)}\right) \nonumber\\
&\equiv& \frac{a_0}{\sqrt{\alpha'}} R^{eff} \left|\frac{n_w R}{\omega_l a_0 \rho}\right| A\left(\frac{r}{R^{eff}(z)}\right)
\end{eqnarray}
where, $F, A$ are subject to boundary conditions analogous to those imposed before. 
Here $\theta$ may also be interpreted as the angular coordinate of a point on the $S^1$ sub-manifold that defines the effective 
radius of the winding (for any value of $z$), and $r$ is the distance from the ``centre" of the $S^1$ 
to its circumference. 
Physically, a section of wound string effectively occupies a volume $\sim 2\pi R^{eff}(z) dz$ in the large dimensions such that $r$ and $\theta$ 
also admit their usual interpretations of the radial and angular coordinates for a string of finite width.
This gives rise to a flux at each point along the string which is quantised in terms of $n_w$ in the string picture (for fixed values of the other parameters) 
and in terms of $n$ in the field picture
\begin{eqnarray}
\Phi_{n_w}(z) 
= \frac{2\pi a_0}{\sqrt{\alpha'}} R^{eff}(z) \sqrt{\left|\frac{n_w R}{\omega_l a_0 \rho}\right|}
\equiv \Phi_{n}(z) = \frac{2\pi n}{e^{eff}(z)}.
\end{eqnarray}
\subsection{Argument for a time-dependent bead mass}
We now use physical arguments to construct dynamic models of pinch formation, 
which are the direct analogue of dynamical winding formation in the random walk and velocity correlations regimes. 
To construct the analogue of the former; consider a field configuration corresponding to an $|n_1|$-vortex string undergoing random quantum 
fluctuations in both its phase $\theta$ and magnitude $f(r)\eta$ at every point on the string. 
In order for a section of string within the horizon to undergo a spontaneous transition to a new topological state $|n_2| \neq |n_1|$,
the fluctuations in $\theta$ would need to be perfectly correlated over some distance $\Delta z \geq l_p$ (i.e. within a volume of 
approximately $\sim r_c^2 \Delta z$). For example, a transition from a $+|n|$ state to a $-|n|$ state, at even a single position $z$ corresponds 
to degenerate vacuum states $\left\langle \phi\right\rangle = \eta f(r) e^{i|n|\theta}$ ($r\leq r_s,r_v, \ \theta \leq 2\pi$) 
tunnelling through the Mexican hat potential so that $\theta \rightarrow \theta + \delta \theta = \theta + \pi$. This is highly unlikely, 
regardless of the tunnelling amplitude for such a transition at any individual point.
\\
\indent
It is also possible for random fluctuations in the magnitude $\delta(f(r)\eta)$ at each point to result in a total reduction 
of the vortex size to sub-Planckian scale. If this were to occur within the horizon it would be necessary for such fluctuations to be correlated, 
at least over some length scale $\Delta z \geq l_p$, in order for a finite section of string to re-emerge with a different topological winding number. 
Even if this continuously occurs over small regions (which would be the analogue of 
continuous Brownian motion at every point along a wound string), it is clear that the net effect would be to to 
leave the macroscopic structure of the string, and the total number of pinches, unchanged.
\\
\indent
The situation is different, however, for the vortex slice which lies (instantaneously) on the horizon. 
In this case the vortex is free to re-emerge in a differing topological state so that only fluctuations at the horizon may 
contribute to the net number of pinches. 
Let us assume that the edge of the vortex at the horizon moves with velocity $v$, 
but randomly increases or decreases, so that its net velocity may be estimated via
\begin{eqnarray}
\left\langle v\right\rangle \sim \sqrt{v} \sim \frac{n_p r_c}{\sqrt{\rho^2 + n_p^2 r_c^2}}
\end{eqnarray}
where $n_p$ denotes the number of pinches and we have again assumed critical coupling. 
Making the transition to warped space so that $\rho \rightarrow a_0\rho$ and using $R \sim r_c$, the expression for $\left\langle v\right\rangle$ is equivalent to that for $\omega_l$:
\begin{eqnarray}
\left\langle v\right\rangle \sim \sqrt{v} \sim \frac{n_p r_c}{\sqrt{\rho^2 + n_p^2 r_c^2}} \equiv \omega_l \sim \frac{n_w R}{\sqrt{a_0^2 \rho^2 + n_w^2 R^2}}
\end{eqnarray}
The number of pinches per loop, for a loop formed at $t=t_i$ is then 
\begin{eqnarray}
n_p(t_i) \sim \frac{\sqrt{\left\langle v\right\rangle \epsilon_l \alpha t_i}}{r_c}
\end{eqnarray}
where $\epsilon_l$ is again the step length. Setting $\epsilon_l \sim \alpha \eta^{-1} \sim \alpha a_0^{-1}\sqrt{\alpha'}$ 
then results in expressions for $n_p(t_i)$, $\left\langle v\right\rangle(t_i)$ and $\Delta(t_i)$ in the field picture which are the 
exact analogues of those for $n_w(t_i)$, $\omega_l(t_i)$ and $a_0d(t_i)$ in the string picture, namely,
\begin{eqnarray}
n_p &\approx& \frac{1}{\sqrt{2}} \frac{(\alpha t_i)}{r_c}\left(-1 + \sqrt{1+\frac{4}{\eta^2 t_i^2}}\right)^{\frac{1}{2}} \nonumber\\
&\equiv& n_w \approx \frac{a_0}{\sqrt{2}}\frac{(\alpha t_i)}{R}\left(-1 + \sqrt{1+\frac{4\alpha'}{a_0^4 t_i^2}}\right)^{\frac{1}{2}} \nonumber\\
\left\langle v\right\rangle(t_i) &\approx& \frac{1}{2} \eta (\alpha t_i)\left(-1 + \sqrt{1+\frac{4}{\eta^2 t_i^2}}\right) \nonumber\\
&\equiv& \omega_l \approx \frac{1}{2}\frac{a_0^2(\alpha t_i)}{\sqrt{\alpha'}}\left(-1 + \sqrt{1+\frac{4\alpha'}{a_0^4 t_i^2}}\right) \nonumber\\
\Delta(t_i) &\approx& \sqrt{2} r_c \left(-1 + \sqrt{1+\frac{4}{\eta^2 t_i^2}}\right)^{-\frac{1}{2}} \nonumber\\
&\equiv& a_0d(t_i) \approx \sqrt{2}a_0R \left(-1 + \sqrt{1+\frac{4\alpha'}{a_0^4 t_i^2}}\right)^{-\frac{1}{2}}
\end{eqnarray}
where we have also identified $l_p \sim t_p \sim \sqrt{\alpha'}$ and $m_p \sim \sqrt{\alpha'}^{-1}$ and where we have also made the substitution $t_i \longleftrightarrow a_0t_i$ where necessary when transfering between warped and unwarped space. One may check that under these identifications
\begin{eqnarray}
|n|(t_i) \sim \frac{R}{\omega_l(t_i)a_0d(t_i)} \sim \frac{\sqrt{\alpha'}}{a_0 R}|n_w(t_i)|^{-1}
\end{eqnarray}
so that, as $n_w(t_i) \rightarrow \frac{\sqrt{\alpha'}}{a_0R}$ as $t_i \rightarrow \infty$, $|n|(t_i) \rightarrow |n| \sim 1$. Though differing from our previous result $|n| \sim 1/a_0$ this is to be expected for windings formed via a random walk where the proportion of the total string length contained in the extra-dimensional windings tends to zero as $t_i \rightarrow \infty$. 
Importantly we can now predict the vorticity of the field theoretic strings, in terms of the velocity associated with the fluctuations at the horizon 
(or equivalently in terms of the observed number of pinches). If a suitable quantum theory of vortices
is developed, it may be possible to predict both $|n|$ and $\Delta$ (or $n_p$) in terms of the underlying dynamics.
As a first step towards this goal, note that in the scaling regime, 
\begin{eqnarray}
d(t_i) \sim \frac{1}{2}\frac{a_0^2R}{\sqrt{\alpha'}}t_i  
\end{eqnarray}
was used to identify $a_0d(t_i) \longleftrightarrow \xi(t_i) \sim \gamma t_i$, yielding $\gamma \sim \frac{1}{2}\frac{a_0R}{\sqrt{\alpha'}}$ \cite{LakeThomasWard:JHEP2009}. 
However, we have now constructed a more explicit correspondence between a wound string model in warped space and a defect string in unwarped 
space which further allows us to identify 
\begin{eqnarray}
\gamma &\sim& \frac{1}{2}\frac{a_0R}{\sqrt{\alpha'}} \sim \frac{1}{2}\frac{\alpha}{|n_w(t_i \rightarrow \infty)|} \nonumber\\
&\sim& \frac{1}{2} \frac{1}{\sqrt{\lambda}} \sim \frac{1}{2} \frac{\alpha}{|n_p(t_i \rightarrow \infty)|}
\end{eqnarray}
The physical constraints $|n_w(t_i \rightarrow \infty)| \geq 1$ and $\alpha$, $\gamma \leq 1$ then imply
\begin{eqnarray}
\lambda \sim e^2/|n| \geq 1, 
\end{eqnarray}
which in conjunction with (\ref{eq:Izzy}) yield $Mg_s \leq a_0^{-2}$, though this is perfectly consistent with the SUGRA approximation $Mg_s >> 1$ for $a_0^2 << 1$.
Recall that the total mass of a pinched string loop is 
\begin{eqnarray}
M_T \sim 2\pi \eta^2 |n| \rho + 2\pi \eta^2 |n| \times \frac{1}{2} \frac{r_c^2}{\Delta^2} \int^{z=n_p\Delta}_{z=0}|G'|^2 dz,
\end{eqnarray}   
which, for the ansatz (\ref{eq:G_ansatz}), may in turn may be approximated by
\begin{eqnarray}
M_T \sim 2\pi \eta^2 |n| \rho + 2\pi \eta^2 |n| \times \frac{\pi}{2} \frac{r_c^2}{\Delta^2} \times n_p\Delta.
\end{eqnarray}  
Although further investigation is needed, we content ourselves here with demonstrating an order of magnitude equivalence 
between the pinched string bead mass and the wound string bead mass. In terms of field-theoretic parameters, the bead mass is then
\begin{eqnarray} \label{eq:Bazinga}
M_b \sim \frac{\pi}{4} \eta^2 |n| \times \frac{r_c^2 n_p}{\Delta}  
\end{eqnarray} 
where we have used the fact that the mass of an individual bead is half that associated with a single pinch. 
We then make use of the recently established identities
$\eta \sim a_0 \sqrt{\alpha'}^{-1}$, $T_1 \sim T_{(1,0)} \sim {\alpha'}^{-1}$, $r_c \sim R$, $n_p \sim n_w$ and $a_0 d \sim \Delta$ to obtain
\begin{eqnarray}
M_b \sim \frac{\pi}{4} T_1 \frac{a_0 n_w R^2}{d} |n|. 
\end{eqnarray} 
Using $a_0d \sim \frac{a_0 \rho}{n_w}$ we then find
\begin{eqnarray}
M_b \sim \frac{\pi}{4} T_1 \frac{n_w^2 R^2}{\rho} a_0|n|, 
\end{eqnarray}
which is in turn equivalent to,
\begin{eqnarray} \label{eq:OMG}
M_b \sim \frac{\pi}{4} T_1 \frac{n_w^2 R^2}{\rho}
\end{eqnarray}
for $|n| \sim 1/a_0$. This result is identical to that obtained in \cite{LakeThomasWard:JCAP2010} for the mass of a wound-string bead in which the windings were formed via a random walk regime. However, as mentioned previously, in this case we expect that $|n| \sim 1$, rather than $|n|\sim1/a_0$ (which we expect to associate with windings formed in a velocity correlations regime), which would instead yield 
\begin{eqnarray} \label{eq:OMG2}
M_b \sim \frac{\pi}{4} T_1 \frac{n_w^2 R^2a_0}{\rho}.
\end{eqnarray}
There therefore remains a slight inconsistency between the result presented here and those obtained in the previous paper, though it may be hoped that further analysis may be able to clarify this point. The most important result is that, in either case (i.e. whether the field-theory expression (\ref{eq:Bazinga}) leads to (\ref{eq:OMG}) or (\ref{eq:OMG2}) in the string picture), the late time fall-off of $M_b(t_i)$ is proportional to $t_i^{-1}$, and the behaviour of pinched string networks may closely mimic that of necklaces formed from wound strings. 
\\
\indent
Additionally, we may hope to use the dual pinched string model to answer the question: does bead mass remain fixed after the 
time of loop formation $t_i$? Unfortunately, in the absence of a full dynamical model of pinch evolution, we are still unable to answer this question 
with any certainty. 
As the bead mass in this model depends on the quantity $\left(\frac{dr_c^{eff}}{dz}\right)^2 \sim \frac{r_c^2}{\Delta^2}(G')^2$ 
we see that two competing factors come into play - the time evolution of $\Delta, (\Delta(t_i,t)$ in an explicitly dynamic model) 
and that of the gradient term $(G')^2$. In principle it is possible for either factor to outweigh the other, 
so that the bead mass could even increase! This too may have an analogue in the string picture, whereby the contraction of the string causes neighbouring 
windings to move closer together, increasing their effective radii. Much more work is needed in both the string and field theory 
models in order to develop a full dynamical theory.
\\
\indent
Finally, we make a physical observation regarding the construction of a model of pinch formation that is analogous to winding production in the velocity correlations regime. In this case we assume that the edge of the vortex at the horizon has a classical velocity 
causing the string core to shrink to the Planck scale before growing again to its maximum radius (reversing direction). 
Dynamically it is unclear why this should be so, but an alternate view of this scenario is that the 
defect string formed ready-pinched at the epoch of the symmetry-breaking phase transition, 
so that successive pinches (separated by a characteristic length scale $\Delta$) are then simply uncovered by the advancing horizon at a rate proportional to $t$.
%
\section{Conclusions}

We have obtained static vortex solutions to a modified Abelian-Higgs model which describe non-cylindrical strings by introducing spatially-dependent couplings. Hypothesising the existence of Planck-sized regions in which vorticity becomes undefined, 
it was also possible to construct a model in which neighbouring sections of string carry different topological charges.
Assuming a periodic variation in the pinched string profile, a formal correspondence between the resulting periodic tension and the 
effective four-dimensional tension of a wound $F$/$D$-string in the KS geometry was obtained. 
Using a specific, but natural, ansatz choice for the string embedding describing non-geodesic windings, 
we were able to obtain specific identities between string theory parameters (defining the KS geometry) and the field theory parameters (defining the Abelian-Higgs model). In the dual string picture, the spatial dependence of the field couplings were found to be related to the effective radius of the windings.
\\
\indent
One interesting observation about this matching is that the field theory result is at large $|n|$ (i.e. large winding number), whilst
the $(p,q)$-string tension has large flux \cite{Thomas:2006ud, Firouzjahi:2006xa}. More concretely the matching on the field theory side was through the parameter pairing $|n| \eta^2$, with $a_0 T_1$ on the string theory side, which suggests a relation between the symmetry breaking scale (set by $\eta$) and
the string length.
\\
\indent
Though there are many possible ways in which to improve and build upon this work, the most valuable 
would be to extend the present analysis from the purely static case to the more general dynamical one. 
Ideally complete dynamical models of pinched string formation (and pinch evolution) would be developed which would allow us to 
determine the cosmological consequences of pinched string networks with greater accuracy. 
It would also be instructive to compare these with general dynamical models of wound $(p,q)$-strings in order to determine if correspondences 
exist for more general string backgrounds. A simple example would be the inclusion
of fluxes present in the warped deformed conifold solution, but analysis in more general backgrounds would 
be most likely to lead to a better understanding of the field-string correspondence.
Moreover we have only considered models of string necklaces in this paper, and the extension to more
general string lattice configurations (where each bead is attached to $N > 2$ strings) remains to be developed \cite{Martins:2010ma, Martins:2008ks}
There is significant scope for future work, including the matching of such models to the CMB anisotropy data and evaluating 
the projected gravitational wave spectrum. We intend to return to such topics in future. 
The interested reader is referred to \cite{LakeThesis} for a more detailed discussion of these topics.

\begin{center}
{\bf Acknowledgments}
\end{center}
We wish to thank V. M. Red'kov, Yuri Sitenko and S. Thomas for their insightful questions and comments. JW is supported
by NSERC of Canada.

\end{document}